\documentclass[aps,prb,twocolumn,noshowpacs,citeautoscript,10pt,amsmath,amssymb,amsfonts,floatfix]{revtex4-1}

\usepackage{times}

\usepackage{bm}

\usepackage{graphicx}

\newcommand{\unitvec}[1]{\hat{\mathbf{#1}}}
\def\doo{\partial}
\def\im{i}

\begin{document}

\title{Half-quantum vortices and walls bounded by strings in the polar-distorted phases of topological superfluid $^3$He}

\author{J.T.~M\"akinen$^{1\ast}$}
\author{V.V.~Dmitriev$^{2}$}
\author{J.~Nissinen$^{1}$}
\author{J.~Rysti$^{1}$}
\author{G.E.~Volovik$^{1,3}$}
\author{A.N.~Yudin$^{2}$}
\author{K. Zhang$^{1,4}$}
\author{V.B.~Eltsov$^{1}$}

\affiliation{$^{1}$Low Temperature Laboratory, Department of Applied Physics, Aalto University, FI-00076 AALTO, Finland; *E-mail:   jere.makinen@aalto.fi \\
$^{2}$P. L. Kapitza Institute for Physical Problems of RAS, 119334 Moscow, Russia \\
$^{3}$Landau Institute for Theoretical Physics, 142432 Chernogolovka, Russia.\\
$^{4}$University of Helsinki, Department of Mathematics and Statistics, P.O. Box 68 FIN-00014, Helsinki, Finland}


\date{\today}

%

\maketitle


\textbf{
Symmetries of the physical world have guided formulation of fundamental laws, including relativistic quantum field theory and understanding of possible states of matter. Topological defects (TDs) often control the universal behavior of macroscopic quantum systems, while topology and broken symmetries determine allowed TDs. Taking advantage of the symmetry-breaking patterns in the phase diagram of nanoconfined superfluid $^3$He, we show that half-quantum vortices (HQVs) -- linear topological defects carrying half quantum of circulation -- survive transitions from the polar phase to other superfluid phases with polar distortion. In the polar-distorted A phase, HQV cores in 2D systems should harbor non-Abelian Majorana modes. In the polar-distorted B phase, HQVs form composite defects -- walls bounded by strings hypothesized decades ago in cosmology. Our experiments establish the superfluid phases of $^3$He in nanostructured confinement as a promising topological media for further investigations ranging from topological quantum computing to cosmology and grand unification scenarios.
}

\vspace{0.1cm}

TDs generally form in any symmetry-breaking phase transitions. The exact nature of the resulting TDs depends on the symmetries before and after the transition. Our universe has undergone several such phase transitions after the Big Bang. As a consequence, a variety of TDs might have formed during the early evolution of the Universe, where phase transitions lead to unavoidable defect formation via the Kibble-Zurek mechanism \cite{Kibble1976,Zurek1985}. Experimentally accessible energy scales $\lesssim 1$~TeV are currently limited to times $t \gtrsim 10^{-12}$~s after the Big Bang by the Large Hadron Collider. Theoretical understanding may be extended up to the Grand Unification energy scales $\lesssim 10^{15}$~GeV of the electroweak and strong forces ($t \gtrsim 10^{-36}\dots 10^{-32}$~s). The nature of the interactions before this epoch remains unknown \cite{SuSy1,RevModPhys.72.25}, but yet unobserved cosmic TDs, the nature of which depends on the Grand Unified Theory (GUT) in question, may help us limit the possibilities. Predictions exist for point defects, such as the t'Hooft-Polyakov magnetic monopole \cite{1974NuPhB..79..276T,Polyakov1974}, linear defects or strings \cite{Kibble1976}, surface defects or domain walls \cite{Zeldovich1974}, and three-dimensional textures \cite{Cruz1612}.

Even though cosmic TDs have not been detected, many of their condensed-matter analogs have been reproduced in the laboratory, where they have an enormous impact on the behavior of the materials they reside in \cite{doi:10.1146/annurev-conmatphys-031016-025154}. Examples include vortices in superconductors \cite{PhysRevLett.62.214}, vortices and monopoles in ultracold gases \cite{Zwierlein2005,Mottonen2014}, and skyrmions in chiral magnets \cite{Muhlbauer915}. Superfluid phases of $^3$He offer an experimentally accessible system to
\begin{figure}
\includegraphics[width=0.9\linewidth]{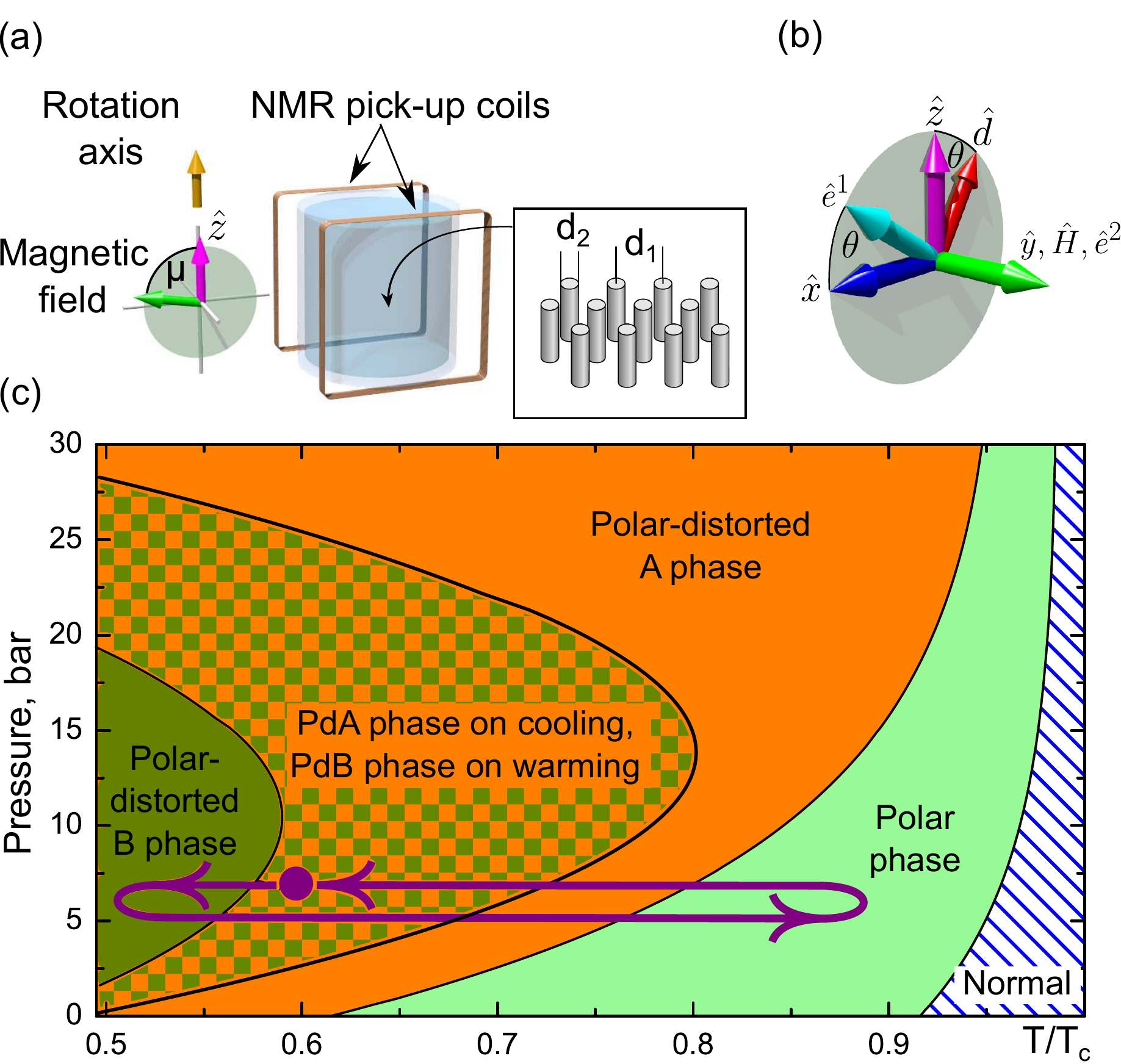}
\caption{{\bf The experimental setup and superfluid phase diagram in nanoconfinement.} (a) The $^3$He sample is confined within a cylindrical container filled with commercially available nanomaterial called nafen-90 (where the number refers to its density in mg/cm$^3$) with uniaxial anisotropy, which consists of nearly parallel Al$_2$O$_3$ strands with $d_{2} \approx 8$~nm diameter, separated by $d_{1} \approx 50$~nm on average. The strands are oriented predominantly along the axis denoted as $\hat{\mathbf{z}}$. The sample can be rotated with angular velocities up to 3~rad/s around the same axis $\hat{\mathbf{z}}$. The sample is surrounded by rectangular nuclear magnetic resonance (NMR) pick-up coils. The static magnetic field transverse to the NMR coils can be oriented at an arbitrary angle $\mu$ with respect to the $\hat{\mathbf{z}}$ axis. (b) The magnetic field, oriented along the $y$-direction ($\mu = \pi/2$) in this figure, locks the $\hat{\mathbf{e}}^2$-vector in the polar-distorted B phase order parameter, Eq.~\eqref{distBop_main}. Vectors $\hat{\mathbf{d}}$ and $\hat{\mathbf{e}}^1$ are free to rotate in the $xz$-plane by angle $\theta$. (c) Sketch of the superfluid phase diagram in our sample in units of $T_{\mathrm{c}}$ of the bulk fluid \cite{PhysRevLett.115.165304}. The purple arrows illustrate the thermal cycling used in the measurements and the purple marker shows a typical measurement point within the region where either polar-distorted phase can exist, depending on the direction of the temperature sweep. The thermal cycling is performed at constant 7 bar pressure.}
\label{CellFig}
\end{figure}
study a variety of TDs and the consequences of symmetry-breaking patterns owing to its rich order-parameter structure resulting from the $p$-wave pairing. Analogs of exotic TDs, such as the Witten string \cite{Witten:1984eb} -- the broken-symmetry-core vortex in superfluid $^3$He-B \cite{PhysRevLett.67.81,Thuneberg2015,Volovik1990}, the skyrmion texture in superfluid $^3$He-A \cite{RevModPhys.59.533}, and the Alice string \cite{SCHWARZ1982141} -- half-quantum vortex (HQV) in the polar phase of superfluid $^3$He \cite{Autti2016}, have been observed.

Here we focus on composite defects -- combinations of TDs and/or non-topological defects of different dimensionality \cite{0954-3899-42-9-094002,Kibble1982a,PhysRevLett.85.4739}. Such defects appear in some GUTs and even in the Standard Model, where the Nambu monopole may terminate an electroweak string \cite{NAMBU1977505,ACHUCARRO2000347}. There are two mechanisms for the formation of composite defects: the hierarchy of energy/interaction length scales \cite{MineevVolovik1978,PhysRevLett.85.4739,Kibble2000}, and the hierarchy (sequential order) of the symmetry-breaking phase transitions \cite{Kibble1982a,Kibble1982b}. 
Composite defects originating from the hierarchy of length scales  of condensation, magnetic, and spin-orbit energies are well-known in superfluid $^3$He. For example, the spin-mass vortex in $^{3}$He-B \cite{Kondo1992,PhysRevLett.85.4739} has a hard core of the coherence length size, defined by the condensation energy, and a soliton tail with thickness of the much larger spin-orbit length. A half-quantum vortex (HQV) originally predicted to exist in the chiral superfluid $^3$He-A \cite{VolovikMineev1976} has a similar structure with the soliton tail, which makes these objects energetically unfavorable.

Composite defects related to the hierarchy of symmetry-breaking phase transitions were discussed in the context of the GUT scenarios by Kibble, Lazarides, and Shafi \cite{Kibble1982a,Kibble1982b}. Here the GUT symmetry, such as $Spin(10)$, is broken into the Pati-Salam group $SU(4)\times SU(2)\times SU(2)$, which in turn is broken to the Standard Model symmetry group $SU(3)\times SU(2)\times U(1)$. At the first transition the linear defects -- cosmic strings,  become topologically stable, while after the second transition they are no longer supported by topology and form the boundaries of the nontopological domain walls, henceforth referred to as Kibble-Lazarides-Shafi (KLS) walls. To the best of our knowledge, observations of KLS walls bounded by strings have not been reported previously.

In this work we explore experimentally the composite defects formed by both the hierarchy of energy scales and the hierarchy of symmetry-breaking phase transitions allowed by the phase diagram of superfluid $^3$He confined in nematically ordered aerogel-like material called nafen. In our sample a sequence of the polar, chiral polar-distorted A (PdA) and fully gapped polar-distorted B (PdB) phases occurs on cooling from the normal state \cite{PhysRevLett.115.165304}, see Fig.~\ref{CellFig}~(c). Previously we established a procedure to form topologically protected HQVs in the polar phase \cite{Autti2016}. At the transition from the polar phase to the PdA phase we expect the HQVs to acquire spin-soliton tails with the width of the spin-orbit length which is much larger than the coherence-length size of vortex cores. On a subsequent transition to the PdB phase, the symmetry breaks in such a way that HQVs lose topological protection and may exist only as boundaries of the non-topological KLS walls. Simultaneously, the spin solitons between HQVs are preserved in the PdB phase and such an object becomes a doubly-composite defect. Naively, however, one would expect that a much stronger tension of the KLS wall compared to that of the spin soliton, would lead to collapse of an HQV pair, possibly to a singly quantized vortex with an asymmetric core \cite{PhysRevLett.67.81,Thuneberg2015,Volovik1990,SalomaaVolovik1988,Thuneberg2014}.

Here we report evidence that HQVs do exists in the superfluid PdA and PdB phases of $^3$He.  We create an array of HQVs by rotating the container with the angular velocity $\Omega$ in zero magnetic field during the transition from the normal fluid to the polar phase \cite{Autti2016} and proceed by cooling the sample through consecutive transitions to the PdA and PdB phases. The HQVs are identified based on their NMR signature as a function of temperature and $\Omega$. A characteristic satellite peak present in the NMR spectrum confirms that the HQVs survive in the PdA phase, where they provide experimental access to vortex-core-bound Majorana states \cite{PhysRevB.61.10267,Volovik1999}. Moreover, the HQVs are found to survive the transition to the PdB phase. The observed features of the NMR spectrum in the PdB phase suggest that a KLS wall emerges between a pair of HQVs already connected by the spin soliton. Evidently the tension of the KLS wall is not sufficient to overcome the pinning of HQVs in nafen. Vortex pinning allows us to study the properties of the out-of-equilibrium vortex state created during the superfluid phase transitions while suppressing the vortex dynamics. Simultaneously pinning does not affect the symmetry-breaking pattern leading to formation of the KLS walls. Our results show that pinned TDs, once created, may be transferred to new phases of matter with engineered topology \cite{Levitin841,PhysRevB.92.144515,polar_TO}.


\section*{Results}

The superfluid phase diagram under confinement by nafen \cite{PhysRevLett.115.165304} -- a nanostructured material consisting of nearly parallel strands made of Al$_2$O$_3$, c.f. Fig.~\ref{CellFig}~(b) -- differs from that of the bulk $^3$He; the critical temperature is suppressed and, more importantly, new superfluid phases - the polar, polar-distorted A (PdA), and polar-distorted B (PdB) phases - are observed. We refer to the Supplementary Note 1 for a detailed discussion on these phases and their symmetries and focus on our observations regarding the HQVs in the PdA and PdB phases.

\begin{figure}[t]
\includegraphics[width=\linewidth]{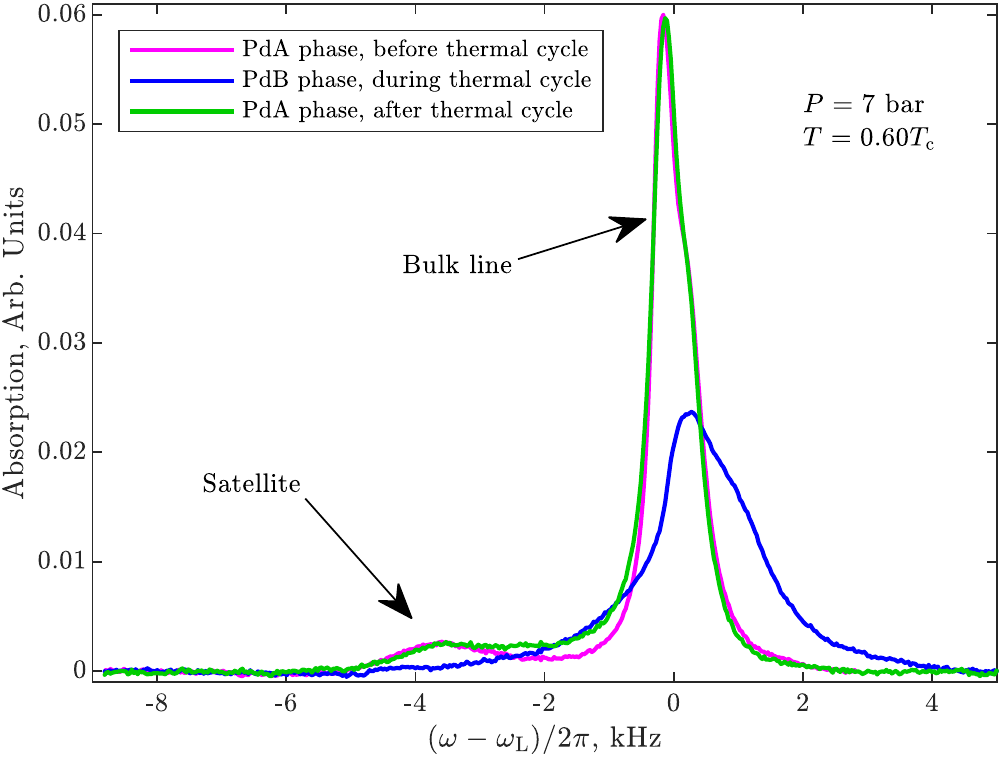}
 \caption{{\bf Survival of HQVs during phase transitions.} The plot shows the measured NMR spectra in transverse ($\mu=\pi/2$) magnetic field in the presence of HQVs. HQVs were created by rotation with 2.5~rad/s during the transition from normal phase to the polar phase. The NMR spectrum includes the response of the bulk liquid and the $\hat{\mathbf{d}}$-solitons, which appear as a characteristic satellite peak at lower frequency. The satellite intensity in the PdA phase remains unchanged after thermal cycling presented in Fig.~\ref{CellFig}~(c). The NMR spectrum in the PdB phase at the same temperature, measured between the two measurements in the PdA phase, is shown for reference.} 
 \label{dist_A_spectra_before_and_after}
\end{figure}

\subsection*{Half-quantum vortices in the PdA phase}

The order parameter of the PdA phase can be written as
\begin{equation}
 A_{\alpha j} = \sqrt{\frac{1+b^2}{3}} \Delta_{\mathrm{PdA}} e^{i \phi} \hat{ \mathbf{d}}_\alpha (\hat{\mathbf{m}}_{j} + ib\hat{\mathbf{n}}_{j}),
\end{equation}
where the orbital anisotropy vectors $\hat{\mathbf{m}}$ and $\hat{\mathbf{n}}$ form an orthogonal triad with the Cooper pair orbital angular momentum axis $\hat{\mathbf{l}} = \hat{\mathbf{m}} \times \hat{\mathbf{n}}$, and $\hat{\mathbf{d}}$ is the spin anisotropy vector. Vector $\hat{\mathbf{m}}$ is fixed parallel to the nafen strands. The amount of polar distortion is characterized by a dimensionless parameter $0 < b < 1$ and  $\Delta_{\mathrm{PdA}}(T,b)$ is the the maximum gap in the PdA phase. The order parameter of the polar phase is obtained for $b=0$, while $b=1$ produces the order parameter of the conventional A phase.

In our experiments, we use continuous-wave NMR techniques to probe the sample, see Methods for further details. In the superfluid state the spin-orbit coupling provides a torque acting on the precessing magnetization, which leads to a shift of the resonance from the Larmor value $\omega_{\mathrm{L}} = | \gamma | H,$ where $\gamma = -2.04 \times 10^{8}$~s$^{-1}$T$^{-1}$ is the gyromagnetic ratio of $^3$He. The transverse resonance frequency of the bulk fluid with magnetic field in the direction parallel to the strand orientation, i.e. $\mu=0$ in Fig.~\ref{CellFig}~(a), is \cite{PhysRevLett.115.165304}
\begin{equation} \label{eq:distAmain}
 \Delta \omega_{\mathrm{PdA}} = \omega_{\mathrm{PdA}}-\omega_{\mathrm{L}} \approx \frac{\Omega_{\mathrm{PdA}}^{2}}{2 \omega_{\mathrm{L}}},
\end{equation}
where $\Omega_{\mathrm{PdA}}$ is the frequency of the longitudinal resonance in the PdA phase at $\mu=\pi/2$. The NMR line retains its shape during the second order phase transition from the polar phase but renormalizes the longitudinal resonance frequency due to appearance of the order parameter component with $b$.

Quantized vortices are linear topological defects in the order-parameter field carrying non-zero circulation. In the PdA phase quantized vortices involve phase winding by $\phi \rightarrow \phi+2\pi\nu$ and possibly some winding of the $\hat{\mathbf{d}}$ vector. The typical singly quantized vortices, also known as phase vortices, have $\nu=1$ and no winding of the $\hat{\mathbf{d}}$-vector, while the HQVs have $\nu=\frac{1}{2}$ and winding of the $\hat{\mathbf{d}}$-vector by $\pi$ on a loop around the HQV core so that sign changes of $\hat{\mathbf{d}}$ and of the phase factor $e^{i \phi}$ compensate each other. The reorientation of the $\hat{\mathbf{d}}$-vector leads to the formation of $\hat{\mathbf{d}}$-solitons -- spin-solitons connecting pairs of HQVs. The soft cores of the $\hat{\mathbf{d}}$-solitons provide trapping potential for standing spin waves \cite{VollhardtWoelfle}.

Since the $\hat{\mathbf{m}}$-vector is fixed by nafen parallel to the anisotropy axis, the $\hat{\mathbf{l}}$-vector lies on the plane perpendicular to it, prohibiting the formation of continuous vorticity\cite{PhysRevLett.36.594} like the double-quantum vortex in $^3$He-A \cite{doublequantum}. Some planar structures in the $\hat{\mathbf{l}}$-vector field, such as domain walls \cite{PhysRevLett.120.205301} or disclinations, remain possible but the effect of the $\hat{\mathbf{l}}$-texture on the trapping potential for spin waves is negligible due to the large polar distortion \cite{PhysRevLett.115.165304} (i.e. for $b \ll 1$). Recent theoretical work \cite{hqv_stability} provides arguments why formation of HQVs in the polar phase is preferred compared to the A phase. Indeed, in confined geometry where the PdA phase is observed immediately below $T_{\mathrm{c}}$, no HQVs are found \cite{PhysRevLett.120.075301}. In our case the PdA phase is obtained via the second-order phase transition from the polar phase with preformed HQVs. We already know \cite{Autti2016} that the maximum tension from the spin-soliton in the polar phase (for $\mu = \pi/2$) is insufficient to overcome HQV pinning. Thus, survival of HQVs in the PdA phase is expected. Moreover, we note that even for $|b|=1$ and in the absence of pinning, a pair of HQVs, once created, should remain stable with finite equilibrium distance corresponding to cancellation of vortex repulsion and tension from the soliton tail \cite{RevModPhys.59.533}.

\begin{figure*}
 \includegraphics[width=\linewidth]{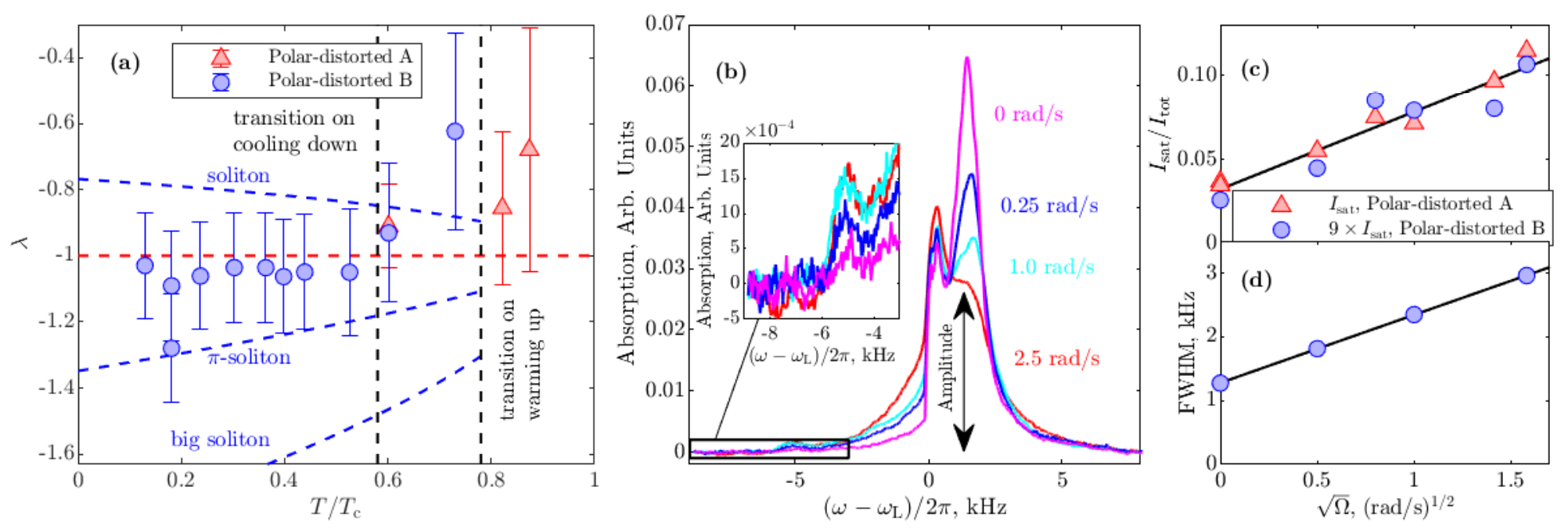}
\caption{{\bf NMR spectra and spin-solitons in the polar-distorted phases.} (a) Frequency shift of a characteristic satellite peak in the NMR spectrum expressed via parameter $\lambda$ as a function of temperature in the PdA and PdB phases. In the PdA phase the measured values reside slightly above the theoretical prediction for a $\hat{\mathbf d}$-soliton with $\pi$ winding, shown as the red dashed line. The difference is believed to be caused by disorder introduced by nafen, as in the polar phase \cite{Autti2016,spinglass}. The corresponding values in the PdB phase for the lowest-energy $\hat{\mathbf{d}}$-soliton (marked ``soliton'') and its antisoliton (marked ``big soliton''), as well as the combined $\pi$-soliton (see text) are shown as dashed blue lines. The $\pi$-soliton values turn out to be in the same ratio with respect to the experimental points as in the PdA phase. The error bars denote the uncertainty in the position of the satellite peak by 1.0~kHz and 0.5~kHz in the PdB and PdA phases, respectively. The uncertainty is taken as the full width at half maximum (FWHM) of the satellite peak in the PdB phase and as half of the FWHM due to improved signal-to-noise ratio in the PdA phase. (b) The plot shows the measured NMR spectrum in the PdB phase at 0.38~$T_{\mathrm{c}}$ for different HQV densities, controlled by the angular velocity $\Omega$ at the time of crossing the $T_{\rm c}$. The presence of  KLS  walls produces characteristic features seen both as widening of the main line (with small positive frequency shift) and as a satellite peak with a characteristic negative frequency shift. The inset shows magnified view of the satellite peak. (c) The satellite intensity in the PdA phase at $0.60 T_{\mathrm{c}}$ (blue circles) and in the PdB phase multiplied by a factor of 9 (red triangles) at $0.38 T_{\mathrm{c}}$ show the expected $\sqrt{\Omega}$-scaling. The solid black line is a linear fit to the measurements including data from both phases. The non-zero $\Omega=0$ intersection corresponds to vortices created by the Kibble-Zurek mechanism \cite{Kibble1976,Zurek1985,Autti2016}. (d) The FWHM of the main line, determined from the spectrum in panel (b), gives FWHM $\approx 3$~kHz for 2.5~rad/s. FWHM for other angular velocities is recalculated from the amplitude of the main NMR line, shown in panel (b), assuming constant area.} 
 \label{satellites_combined}
\end{figure*}

In the presence of HQVs the excitation of standing spin waves localized on the soliton leads to a characteristic NMR satellite peak in transverse ($\mu=\pi/2$) magnetic field, c.f. Fig.~\ref{dist_A_spectra_before_and_after}, with frequency shift
\begin{equation} \label{eq:lambda}
 \Delta \omega_{\mathrm{PdAsat}} = \omega_{\mathrm{PdAsat}} - \omega_{\mathrm{L}} \approx \lambda_{\mathrm{PdA}} \frac{\Omega_{\mathrm{PdA}}^{2}}{2 \omega_{\mathrm{L}}},
\end{equation}
where $\lambda_{\mathrm{PdA}}$ is a dimensionless parameter dependent on the spatial profile (texture) of the order parameter across the soliton. For an infinite 1D $\hat{\mathbf{d}}$-soliton, one has $\lambda_{\mathrm{PdA}} = -1$, corresponding to the zero-mode of the soliton \cite{Autti2016,RevModPhys.59.533,HuMaki87}. The measurements in the supercooled PdA phase, Fig.~\ref{satellites_combined} (a), at temperatures close to the transition to the PdB phase give value $\lambda_{\mathrm{PdA}} \approx -0.9$, which is in good agreement with theoretical predictions and earlier measurements in the polar phase with a different sample \cite{Autti2016}. This confirms that the structure of the $\hat{\mathbf{d}}$-solitons connecting the HQVs is similar in polar and PdA phases and the effect of the orbital part to the trapping potential can safely be neglected. Detailed analysis of the satellite frequency shift as a function of magnetic field direction in the PdA phase remains a task for the future.

\subsection*{Half-quantum vortices in the PdB phase}

Since the HQVs are found both in the polar and PdA phases, a natural question is to ask what is their fate in the PdB phase? The number of HQVs in the polar and PdA phases can be estimated from the intensity (integrated area) of the NMR satellite, a direct measure of the total volume occupied by the $\hat{\mathbf{d}}$-solitons \cite{Autti2016}. When cooling down to the PdB phase from the PdA phase, one naively expects the HQVs and the related NMR satellite to disappear since isolated HQVs cease to be protected by topology in the PdB phase. However, the measured satellite intensity in the PdA phase before and after visiting the PdB phase remained unchanged, c.f. Fig.~\ref{dist_A_spectra_before_and_after}, which is a strong evidence in favor of the survival of HQVs in the phase transition to the PdB phase. Theoretically it is possible that HQVs survive in the PdB phase as pairs connected by domain walls, i.e. as walls bounded by strings \cite{Kibble1982a}. For very short separation between HQVs in a pair and ignoring the order-parameter distortion by confinement, such construction may resemble the broken-symmetry-core single-quantum vortex of the B phase \cite{Thuneberg2015}. In our case, however, the HQV separation in a pair exceeds the core size by three orders of magnitude. Let us now consider this composite defect in more detail.

The order parameter of the PdB phase can be written as
\begin{equation}
A_{\alpha j} = \sqrt{\frac{1+2q^2}{3}} \Delta_{\mathrm{PdB}} e^{i\phi} (\hat{\mathbf{d}}_\alpha \hat{\mathbf{z}}_{j} + q_{1} \hat{\mathbf{e}}^1_\alpha \hat{\mathbf{x}}_j 
+ q_{2} \hat{\mathbf{e}}^2_\alpha \hat{\mathbf{y}}_j ) \,,
\label{distBop_main}
\end{equation}
where $\lvert q_1 \rvert , \lvert q_2 \rvert \in (0,1) $, $\lvert q_1 \rvert = \lvert q_2 \rvert \equiv q$ describes the relative gap size in the plane perpendicular to the nafen strands, $\hat{\mathbf{e}}^1$ and $\hat{\mathbf{e}}^2$ are unit vectors in spin-space forming an orthogonal triad with $\hat{\mathbf{d}}$, and $\Delta_{\mathrm{PdB}}(T,q)$ is the maximum gap in the PdB phase. For $q=0$ one obtains the order parameter of the polar phase, while $q=1$ recovers the order parameter of the isotropic B phase. We extract the value for the distortion factor, $q \sim 0.15$ at the lowest temperatures from the NMR spectra using the method described in Ref.~\citenum{DistB}, see Supplementary Note 7 for the measurements of $q$ in the full temperature range.

In transverse magnetic field $\mathbf{H}$ exceeding the dipolar field, the vector $\hat{\mathbf{e}}^2$ becomes locked along the field, while vectors $\hat{\mathbf{d}}$ and $\hat{\mathbf{e}}^1$ are free to rotate around the axis $\hat{\mathbf{y}}$, directed along $\mathbf{H}$, with the angle $\theta$ between $\hat{\mathbf{d}}$ and $\hat{\mathbf{z}}$, c.f. Fig.~\ref{CellFig}~(b). The order parameter of the PdB phase in the vicinity of an HQV pair has the following properties. The phase $\phi$ around the HQV core changes by $\pi$ and the angle $\theta$ (and thus vectors $\hat{\mathbf{d}}$ and $\hat{\mathbf{e}}^1$) winds by $\pi$. Consequently, there is a phase jump $\phi \rightarrow \phi + \pi$ and related sign flips of vectors $\hat{\mathbf{d}}$ and $\hat{\mathbf{e}}^1$ along some direction in the plane perpendicular to the HQV core. In the presence of order-parameter components with $q>0$, Eq.~\eqref{distBop_main} remains single-valued if, and only if, $q_{2}$ also changes sign. We conclude that the resulting domain wall separates the degenerate states with $q_2 = \pm q$ and together with the bounding HQVs has a structure identical to the domain wall bounded by strings -- the KLS wall -- proposed by Kibble, Lazarides, and Shafi in Refs.~\citenum{Kibble1982a,Kibble1982b}.

\begin{figure*}
\includegraphics[width=\linewidth]{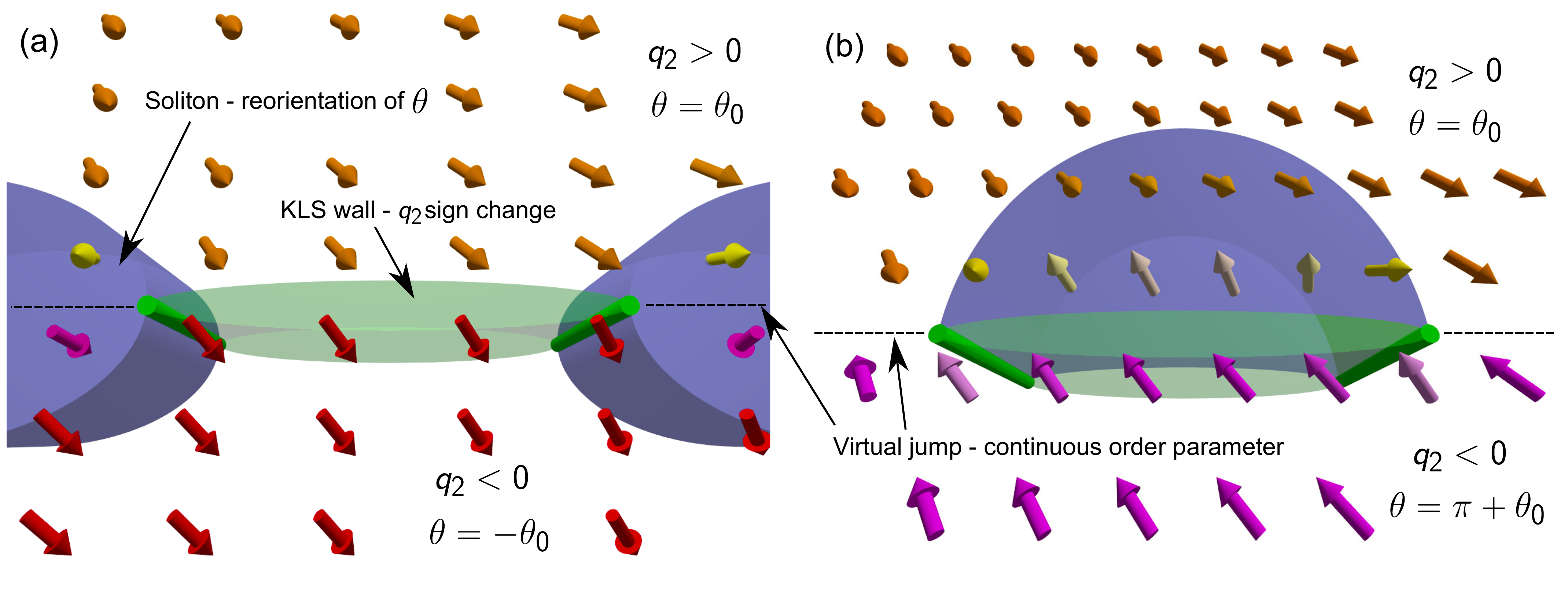}
\caption{{\bf Kibble-Lazarides-Shafi (KLS) wall configurations in the PdB phase.} Each HQV core terminates one soliton - reorientation of the spin part of the order parameter denoted by the angle $\theta$ - and one KLS wall. The orientation of the $\hat{\mathbf{d}}$-vector is shown as arrows where their color indicates the angle $\theta$, based on numerical calculations (Supplementary Note 2). (a) The  KLS  wall is bound between a different pair of HQV cores as the soliton. Ignoring the virtual jumps, the angle $\theta$ winds by $\pi - 2 \theta_{0}$ across the soliton and by $2 \theta_{0}$ across the  KLS  wall. The order parameter is continuous across the virtual jumps, where $\phi \rightarrow \phi + \pi$, $\theta \rightarrow \theta + \pi$, and $q_2 \rightarrow -q_2$. (b) The soliton and the  KLS  wall are bound between the same pair of HQV cores. The total winding of the $\hat{\mathbf{d}}$-vector is $\pi$ across the structure. In principle, the KLS wall may lie inside or outside the soliton. Here the  KLS  wall and the soliton are spatially separated for clarity.}
\label{WallFigCombined}
\end{figure*}

The KLS  wall and the topological soliton  have distinct defining length scales\cite{Volovik1990,Thuneberg2014} -- the KLS wall has a hard core of the order of $\xi_{\mathrm{W}} \equiv q^{-1} \xi$, where $\xi$ is the coherence length, and the soliton has a soft core of the size of the dipole length $\xi_{\mathrm{D}} \gg \xi_{\mathrm{W}}$. The combination of these two objects may emerge in two different configurations illustrated in Fig.~\ref{WallFigCombined}. The minimization of the free energy (Supplementary Notes 3 and 4) shows that in the PdB phase the lowest-energy spin-soliton corresponds to winding of the $\hat{\mathbf{d}}$-vector by $\pi - 2 \theta_{0}$, where $\sin \theta_{0} = q_{2}(2-2q_{1})^{-1}$, on a cycle around an HQV core. Additionally, the presence of KLS walls results in winding of the $\hat{\mathbf{d}}$-vector by $2 \theta_{0}$. These solitons can either extend between different pairs of HQVs, Fig.~\ref{WallFigCombined}~(a), while walls with total change $\Delta \theta = \pi$ are also possible if both solitons are located between the same pair of HQVs, Fig.~\ref{WallFigCombined}~(b).

The appearance of  KLS  walls and the associated $\hat{\mathbf{d}}$-solitons has the following consequences for NMR. The frequency shift of the bulk PdB phase in axial field for $q<1/2$ is \cite{DistB}
\begin{equation} \label{main_shift}
 \Delta \omega_{\mathrm{PdB},\parallel} = \omega_{\mathrm{PdB},\parallel} - \omega_{\mathrm{L}} \approx \left( 1 + \frac{5}{2} q \right)  \frac{\Omega_{\mathrm{PdB}}^{2}}{2 \omega_{\mathrm{L}}},
\end{equation}
where $\Omega_{\mathrm{PdB}}$ is the Leggett frequency of the PdB phase, defined in the Supplementary Note 6. In transverse magnetic field the bulk line has a positive frequency shift
\begin{equation}
 \Delta \omega_{\mathrm{PdB},\perp} = \omega_{\mathrm{PdB},\perp} - \omega_{\mathrm{L}} \approx \left( q - q^2 \right)  \frac{\Omega_{\mathrm{PdB}}^{2}}{2 \omega_{\mathrm{L}}},
\end{equation}
and winding of the $\hat{\mathbf{d}}$-vector in a soliton leads to a characteristic frequency shift
\begin{equation}
 \Delta \omega_{\mathrm{PdBsat}} = \omega_{\mathrm{PdBsat}} - \omega_{\mathrm{L}} \approx \lambda_{\mathrm{PdB}} \frac{\Omega_{\mathrm{PdB}}^{2}}{2 \omega_{\mathrm{L}}},
\end{equation}
where the dimensionless parameter $\lambda_{\mathrm{PdB}}$ is characteristic to the defect. Numerical calculations in a 1D soliton model (Supplementary Note 4) for all possible solitons shown in Fig.~\ref{satellites_combined}~(a) give the low-temperature values $\lambda_{\mathrm{soliton}} \sim -0.8$ for $\pi-2\theta_{0}$-soliton (``soliton'') and $\lambda_{\mathrm{big}} \sim -1.8$ for its antisoliton, which has $\pi + 2\theta_{0}$ winding (``big soliton''). The ($2 \theta_{0}$)-soliton (`` KLS soliton'') related to the  KLS  walls outside spin-solitons gives rise to a frequency shift experimentally indistinguishable from the frequency shift of the bulk line. The last possibility, the ``$\pi$-soliton'' consisting of a KLS soliton and a soliton, c.f. Fig.~\ref{WallFigCombined}~(b), gives $\lambda_\pi \sim -1.3$ at low temperatures. The measured value, $\lambda_{\mathrm{PdB}} \sim -1.1$ at the lowest temperatures, as seen in Fig.~\ref{satellites_combined}~(a). The measured values for $\lambda_{\rm PdB}$, together with the fact that the total winding of the $\hat{\mathbf d}$-vector is also equal to $\pi$ in the PdA and polar phases above the transition temperature, suggest that the observed soliton structure in the PdB phase corresponds to the $\pi$-soliton in the presence of a  KLS  wall.

In addition, the  KLS  wall possesses a tension $\sim \xi q^{3} \Delta_{\mathrm{PdB}} ^{2} N_{0}$ \cite{SalomaaVolovik1988,Thuneberg2014}, where $N_{0}$ is the density of states. Thus the presence of KLS walls applies a force pulling the two HQVs at its ends towards each other. The fact that the number of HQVs remains unchanged in the phase transition signifies that the  KLS  wall tension does not exceed the maximum pinning force in the studied nafen sample. This observation is in agreement with our estimation of relevant forces (see Supplementary Note 5). Strong pinning of single-quantum vortices in B-like phase in silica aerogel has also been observed previously \cite{PhysRevLett.94.075301}. An alternative way to remove a  KLS  wall is to create a hole within it, bounded by a HQV\cite{Kibble1982a}. Creation of such a hole, however, requires overcoming a large energy barrier related to creation of a HQV with hard core of the size of $\xi$. Moreover, growth of the HQV ring is prohibited by the strong pinning by the nafen strands. We also note that for larger values of $q$ there may exist a point at which the KLS wall becomes unstable towards creation of HQV pairs and as a result the HQV pairs bounded by KLS walls would eventually shrink to singly-quantized vortices. For the discussion of the effect of nafen strands on the KLS walls see Supplementary Note 5.

\subsection*{Effect of rotation}

The density of HQVs created in the polar phase is controlled by the angular velocity $\Omega$ of the sample at the time of the phase transition from the normal phase, $n_{\mathrm{\mathrm{HQV}}} = 4\Omega \kappa^{-1}$, where $\kappa$ is the quantum of circulation. The integral of the NMR satellite depends on the total volume occupied by the solitons, whose width is approximately the spin-orbit length  and the height is fixed by the sample size $4$~mm. The average soliton length is equal to the intervortex distance $\propto \Omega^{-1/2}$. Since the number of solitons is half of the number of HQVs, the satellite intensity scales as $\propto \Omega \cdot \Omega^{-1/2} = \sqrt{\Omega}$ which has been previously confirmed by measurements in the polar phase \cite{Autti2016}. Here we observe similar scaling in the PdA and PdB phases, c.f. Fig~\ref{satellites_combined}~(c).

Although the satellite intensity scales with the vortex density in the same way in both phases, there is one striking difference -- the satellite intensity normalized to the total absorption integral in the PdB phase is smaller by a factor of $\sim 9$ relative to the PdA phase. Simultaneously, the original satellite intensity in the PdA phase is restored after a thermal cycle shown in Fig.~\ref{CellFig}~(b). Our numerical calculations of the soliton structure do not indicate that the PdB phase soliton width nor the oscillator strength would decrease substantially to explain the observed reduction in satellite size and the reason for the observed spectral intensity remains unclear -- see Supplementary Note 8 for the calculations.

Another effect of rotation in the PdB phase transverse ($\mu = \pi/2)$ NMR spectrum is observed at the main peak, c.f. Fig.~\ref{satellites_combined}~(b). The full-width-at-half-maximum (FWHM), extracted from the amplitude of the main peak assuming $w\cdot h=\mathrm{const}$, where $w$ is its width and $h$ is height, scales as $\propto \sqrt{\Omega}$; Fig.~\ref{satellites_combined}~(d). Increase in the FWHM may indicate that the presence of  KLS  walls enhances scattering of spin waves and thus results in increased dissipation. Further analysis of this effect is beyond the scope of this Article.

\section*{Discussion}

To summarize, we have found that HQVs, created in the polar phase of $^3$He in a nanostructured material called nafen, survive phase transitions to the PdA and PdB phases. Previously HQVs have been reported in the polar phase \cite{Autti2016}, at the grain boundaries of $d$-wave cuprate superconductors \cite{PhysRevLett.76.1336}, in chiral superconductor rings \cite{Jang186}, and in Bose condensates \cite{Lagoudakis974,PhysRevLett.115.015301}. Of these systems, only the polar phase contains vortex-core-bound fermion states as others are either Bose systems or lack the physical vortex core altogether. The domain walls with the sign change of a single gap component in $^3$He-B were suggested to interpret the experimental observations in bulk samples\cite{Mukharsky2004,Bunkov2006} ($q=1$) and in the slab geometry \cite{stripe_phase}.  Such walls, however, differ from those reported here as they are not bounded by strings but rather terminate at container walls. In the slab geometry such walls are additionally topologically protected by a $\mathbb{Z}_2$ symmetry due to pinning of the $\hat{\mathbf{l}}$ vector by the slab.

The survival of HQVs in the PdA and PdB phases has several important implications. First, HQVs in 2D $p_{\mathrm{x}}+i p_{\mathrm{y}}$ topological superconductors (such as the A or PdA phases) are particularly interesting since their cores have been suggested to harbor non-Abelian Majorana modes, which can be utilized for topological quantum computation \cite{PhysRevLett.86.268}. This fact has attracted considerable interest in practical realization of such states in various candidate systems \cite{Zhang2018,Luthyn2018,RevModPhys.83.1057,doi:10.1146/annurev-conmatphys-030212-184337}. While the PdA phase has the correct $p_{\mathrm{x}} + ip_{\mathrm{y}}$ type order parameter, scaling the sample down to effective 2D remains a challenge for future. However, the presence of the nafen strands, smaller in diameter than the coherence length, increases the separation of the zero-energy Majorana mode from other vortex-core-localized fermion states to a significant fraction of the superfluid energy gap, making it easier to reach relevant temperatures $(k_{\mathrm{B}}T \lesssim$ energy separation of core-bound states$)$ in experiments \cite{PhysRevB.79.134529,0953-8984-25-7-075701}.

Second, we have shown how in the PdB phase the HQVs, although topologically unstable as isolated defects, survive as composite defects known as ``walls bounded by strings'' (here KLS walls bounded by a pair of HQVs) -- first discussed decades ago by Kibble, Lazarides and Shafi in the context of cosmology \cite{Kibble1982a}. Although the present existence of KLS walls in the context of the Standard Model is shown to be unacceptable, as they either dominate the current energy density (first-order phase transition) or disappeared during the early evolution of the Universe (second-order phase transition), they occur in some GUTs and beyond-the-Standard-Model scenarios, especially in ones involving axion dark matter \cite{GPS_DWDM,DM_hunt_with_cloks,Marsh2015}. Any sign of similar defects in cosmological context would thus immediately limit the number of viable GUTs. Under our experimental conditions the transition from the PdA phase to the PdB phase is weakly first-order ($q\ll 1$ at transition), but in principle the order parameter allows a second-order phase transition to the PdB phase directly from the polar phase. Such a phase transition may be realized in future e.g. by tuning confinement parameters. Studying the parameters affecting the amount of supercooling of the metastable PdA state (``false vacuum'') before it collapses to the lowest-energy PdB state (``true vacuum'') may also give insight on the nature of phase transitions in the evolution of the early Universe.

In conclusion, we have shown that the creation and stabilization of HQVs in different superfluid phases with controlled and tunable order parameter structure is possible in the presence of strong pinning by the confinement. The survival of HQVs opens up a wide range of experimental and theoretical avenues ranging from non-Abelian statistics and topological quantum computing to studies of cosmology and GUT extensions of the Standard Model. Additionally, our results pave way for the study of a variety of further problems, such as different fermionic and bosonic excitations living in the HQV cores and within the KLS  walls, and the interplay of topology and disorder provided by the confining matrix \cite{PhysRevB.97.024204}. A fascinating prospect is to stabilize new topological objects possibly in novel superfluid phases by tuning the confinement geometry \cite{Levitin841,PhysRevB.92.144515,polar_TO}, temperature, pressure, magnetic field, or scattering conditions \cite{PhysRevLett.120.075301}.


\section*{Methods}

\subsection*{Sample geometry and thermometry}

The $^3$He sample is confined within a 4-mm-long cylindrical container with $\varnothing$4~mm inner diameter, made from Stycast 1266 epoxy; See Fig.~\ref{CellFig}~(a) for illustration. The experimental volume is connected to another volume of bulk B phase, used for thermometry and coupling to nuclear demagnetization stage. This volume contains a commercial quartz tuning fork with $32$~kHz resonance frequency, commonly used for thermometry in $^3$He \cite{Blaauwgeers2007,PhysRevB.84.224501}. The fork is calibrated close to $T_{\mathrm{c}}$ against NMR signal from bulk $^3$He-B surrounding the nafen-filled volume. At lower temperatures we use a self-calibration scheme \cite{doi:10.1063/1.4891619} by determining the onset of the ballistic regime from the fork's behavior \cite{fork_ballistic}.

\subsection*{Sample preparation}

To avoid paramagnetic solid $^3$He on the surfaces, the sample is preplated with approximately 2.5 atomic layers of $^4$He \cite{PhysRevLett.120.075301}. The HQVs are created by rotating the sample in zero magnetic field with  angular velocity $\Omega$ while cooling the sample from the normal phase to the polar phase. Then the rotation is stopped since, based on our observations, the HQVs remain pinned (and no new HQVs are created) over all relevant time scales, at least for two weeks after stopping the rotation. The typical cooldown rate close to the critical temperature was of the order of $0.01 T_{\mathrm{c}}$ per hour to reduce the amount of vortices created by the Kibble-Zurek mechanism. Once the state had been prepared the temperature was kept below the polar phase critical temperature until the end of the measurement.

\subsection*{NMR spectroscopy}

Static magnetic field of 12--27 mT corresponding to NMR frequencies of 409--841~kHz, is created using two coils oriented along and perpendicular to the axis of rotation. The magnetic field can be oriented at an arbitrary angle in the plane determined by the two main coils. Special gradient coils are used to minimize the field gradients along the directions of the main magnets. The magnetic field inhomogeneity along the rotation axis is $\Delta H_{\mathrm{ax}}/H_{\mathrm{ax}} \sim 10^{-4}$ and in the transverse direction an order of magnitude larger, $\Delta H_{\mathrm{tra}}/H_{\mathrm{tra}} \sim 10^{-3}$. The NMR pick-up coil, oriented perpendicular to both main magnets, is a part of a tuned tank circuit with quality factor $Q \sim 140$. Frequency tuning is provided by a switchable capacitance circuit, thermalized to the mixing chamber of the dilution refrigerator. We use a cold preamplifier, thermalized to a bath of liquid helium, to improve the signal-to-noise ratio in the measurements.

\subsection*{Rotation}

The sample can be rotated about the vertical axis with angular velocities up to $3$~rad/s, and cooled down to $\sim 150 \mu$K using ROTA nuclear demagnetization refrigerator. The refrigerator is well balanced and suspended against vibrational noise. The earth's magnetic field is compensated using two saddle-shaped coils installed around the refrigerator to avoid parasitic heating of the nuclear stage. In rotation, the total heat leak to the sample remains below 20 pW \cite{PhysRevB.84.224501}.

\section*{Acknowledgements}

We thank V.V. Zavyalov and V.P. Mineev for useful discussions and related work on spin-solitons and HQVs. This work has been supported by the European Research Council (ERC) under the European Union's Horizon 2020 research and innovation programme (Grant Agreement No. 694248) and by the Academy of Finland (grants no. 298451 and 318546). The work was carried out in the Low Temperature Laboratory, which is part of the OtaNano research infrastructure of Aalto University.

\section*{Author contributions}

The experiments were conducted by J.T.M. and J.R.; the sample was prepared by V.V.D. and A.N.Y.; the theoretical analysis was carried out by J.T.M., V.V.D., J.N., G.E.V., A.N.Y., and V.B.E.; numerical calculations were performed by J.N. and K.Z.; V.B.E. supervised the project; and the paper was written by J.T.M., J.N., G.E.V., and V.B.E., with contributions from all authors.


%

\clearpage
\newpage

\section*{Supplementary note 1: Symmetries of liquid $^3$He in constrained geometry}

Here we discuss the symmetries possessed by the normal fluid and the superfluid phases under confinement. For schematic illustration of the superfluid gaps in different phases, see Supplementary Figure~\ref{fig:gaps}.

\subsection*{Normal phase}

Above the superfluid transition bulk $^3$He possesses the symmetry group
\begin{equation} \label{eq:bulk_symmetry}
G = SO(3)_{L} \times SO(3)_{S} \times U(1)_{\phi} \times T  \times P
\end{equation}
which includes continuous symmetries: three-dimensional rotations of coordinates $SO(3)_{L}$, rotations of the spin space $SO(3)_{S}$, and the global phase transformation group $U(1)_{\phi}$, as well as discrete symmetries; $T$ is the time-reversal symmetry and $P$ is the space parity symmetry. The transitions from normal fluid to superfluid phases as well as transitions between different superfluid phases are accompanied by the spontaneous breaking of continuous and/or discrete symmetries in $G$ (in addition to the broken $U(1)_{\phi}$ symmetry of the superfluid). In bulk $^3$He three superfluid phases can be realized; the fully-gapped superfluid B phase characterized by broken relative spin-orbit symmetry, the chiral $p_{\mathrm{x}}+ip_{\mathrm{y}}$ state known as the superfluid A phase, and finally, the spin-polarized A$_1$ phase in high magnetic fields.

\begin{figure*}[!t]
 \includegraphics[width=.8\linewidth]{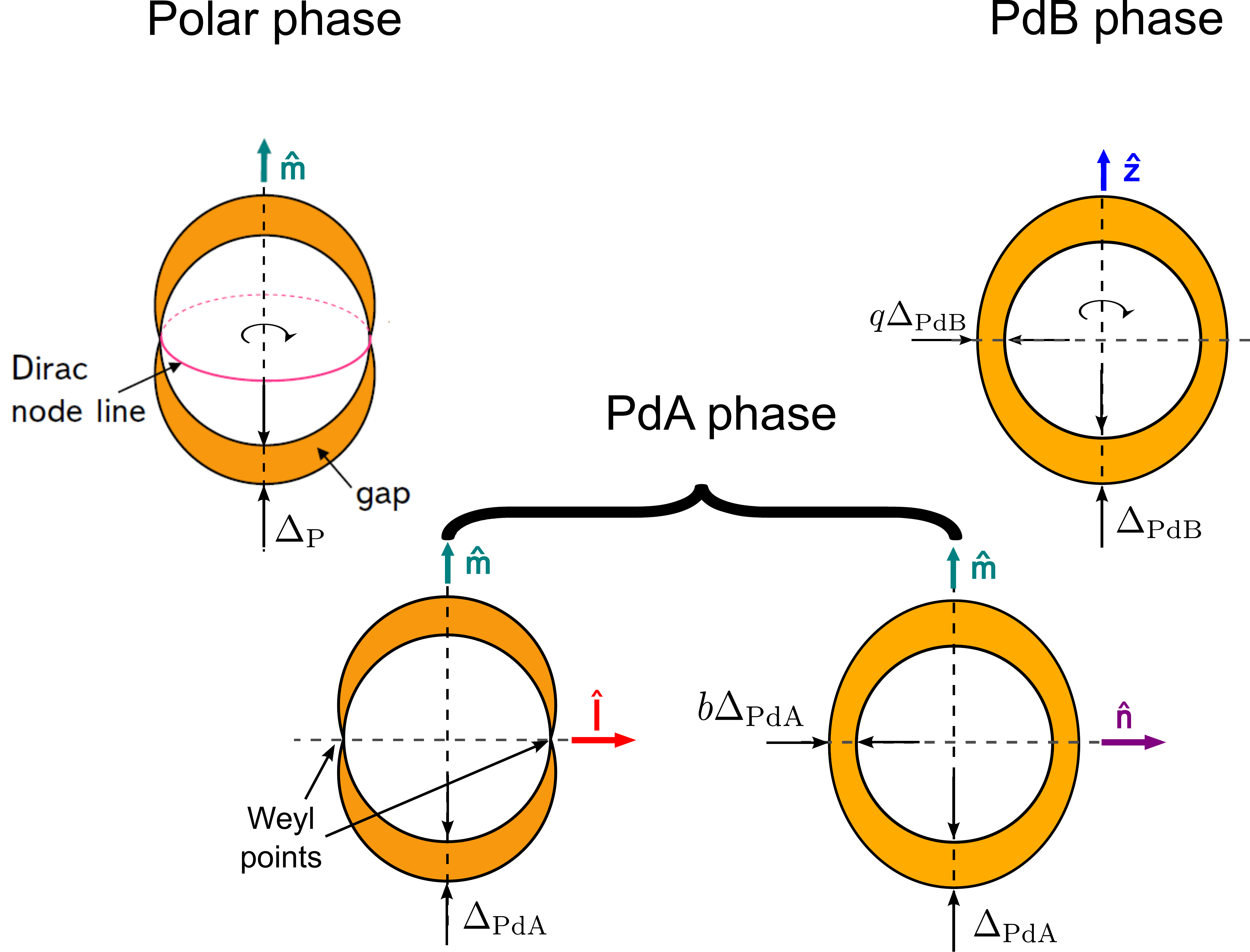}
\caption{The figure shows schematic illustration (not to scale) of superfluid gaps in all superfluid phases encountered under confinement by nafen. The polar phase and PdB phase gaps are symmetric under rotation by the vertical axis, and the PdA phase gap is shown in two projections as it lacks the rotational symmetry.} 
 \label{fig:gaps}
\end{figure*}

In nanostructured confinement, i.e. in thin slabs \cite{3heslab,Levitin841} or in various aerogels \cite{PhysRevLett.112.115303, Wiman2014, PhysRevLett.115.165304,Askhadullin2012,Ikeda2014}, the phase diagram, as well as the symmetry group of the normal phase, can be altered in a controlled fashion. In the presence of commercially available nematically ordered material called nafen \cite{PhysRevLett.115.165304}, the three-dimensional continuous rotational symmetry $SO(3)_{L}$ in Eq.~\eqref{eq:bulk_symmetry} is explicitly broken in the real space by the confinement. As a result, the total symmetry group of the normal phase is reduced to \cite{0953-8984-27-11-113203}
\begin{equation}
 G' = D_{\infty L} \times SO(3)_{S} \times U(1)_{\phi} \times T  \times P \,,
\label{G'group}
\end{equation}
where $D_{\infty L}$ contains rotations about axis $\hat{\bm z}$ and $\pi$ rotations about perpendicular axes. The resulting phase diagram\cite{PhysRevLett.115.165304} differs from that of the bulk $^3$He; the critical temperature is suppressed and, more importantly, new superfluid phases, c.f. Supplementary Figure~\ref{fig:gaps} - the polar, polar-distorted A (PdA), and polar-distorted B (PdB) phases - are observed.

\subsection*{Polar phase}

In our samples, the phase transition with the highest critical temperature always occurs between the normal phase and the polar phase \cite{fomin_theorem}. The order parameter of the polar phase can be written as
\begin{equation} \label{polarop}
 A_{\alpha j} = \frac{1}{\sqrt{3}} \Delta_{\mathrm{P}} e^{i \phi} \hat{\bm{d}}_{\alpha} \hat{\bm{m}}_{j},
\end{equation}
where $\Delta_{\mathrm{P}}(T)$ is the maximum superfluid gap in the polar phase, $\phi$ is the superfluid phase, $\hat{\bm{d}}$ is the unit vector of spin anisotropy, and $\hat{\bm{m}}$ is the unit vector of orbital anisotropy parallel to the anisotropy axis of the confinement. That is, in the transition to the polar phase the orbital part is fixed by the nafen strands and rotational symmetry is preserved only in the plane perpendicular to $\hat{\bm{m}}$. As for any superconducting or superfluid state, the phase acquires an expectation value and the phase gauge symmetry $U(1)_{\phi}$ is broken in the transition. The group describing the remaining symmetries of the polar phase in zero magnetic field is
\begin{equation}
 H_{\mathrm{P}} = \tilde D_{\infty L} \times \tilde D_{\infty S}  \times T \times \tilde P \,.
\end{equation}
Here  the discrete symmetry $\tilde P$ is the inversion $P$ combined with the  phase rotation $e^{\pi i}$. The symmetries $\tilde D_{\infty L}$ and $\tilde D_{\infty S}$ are the symmetries $D_{\infty L}$ and $D_{\infty S}$  in $L$ and $S$ spaces, where the $\pi$ rotations about transverse axes are combined with a phase rotation $e^{\pi i}$. The homotopy group $\pi_1(G'/H_{\mathrm{P}})= \mathbb{Z}\times \mathbb{Z}_2$ provides the topological stability of the phase vortices and the half-quantum vortex. The topological stability of spin vortices is determined by spin-orbit interaction and orientation of the magnetic field \cite{MineevVolovik1978}.

\subsection*{Polar-distorted A phase}

At certain nafen densities and pressures the polar phase transforms on cooling to the polar-distorted A (PdA) phase via a second-order phase transition \cite{PhysRevLett.115.165304}. The order parameter of the PdA phase is
\begin{equation}
 A_{\alpha j} = \sqrt{\frac{1+b^2}{3}} \Delta_{\mathrm{PdA}} e^{i \phi} \hat{\bm{d}}_\alpha (\hat{\bm{m}}_{j} + ib\hat{\bm{n}}_{j}),
\end{equation}
where the vector $\hat{\bm{n}}$ is an orbital anisotropy vector both perpendicular to vector $\hat{\bm{m}}$ and the Cooper pair orbital angular momentum axis $\hat{\bm{l}} = \hat{\bm{m}} \times \hat{\bm{n}}$, and $0 < b < 1$ is a dimensionless parameter characterizing the gap suppression by the confinement. The anisotropy vector $\hat{\bm{l}}$ defines the axis of the Weyl nodes in the PdA phase quasiparticle energy spectrum. The remaining symmetry group in the PdA phase in zero magnetic field is
\begin{equation}
 H_{\rm PdA} =\tilde D_2 \times \tilde D_{\infty S}\times \tilde P \,.
\end{equation}
The time-reversal symmetry is explicitly broken in the PdA phase, while $\tilde P$ combined with $\pi$ orbital rotation about $\hat{\bm{m}}$ remains a symmetry. Together with $\pi$ rotation about the axis $\hat{\bm{m}}\times \hat{\bm{n}}$ combined with the  phase rotation $e^{\pi i}$ the orbital symmetry forms the $\tilde D_2$-group. The group $H_{{\rm PdA}}$ is the subgroup of $H_{{\rm P}}$, which reflects the fact that the PdA phase can be obtained by the second-order phase transition from the polar phase. The homotopy group $\pi_1(G'/H_{{\rm PdA}})= \mathbb{Z}\times \mathbb{Z}\times \mathbb{Z}_2$ provides the topological stability of the phase vortices, the half-quantum vortex and also the orbital disclination in the vector $\hat{\bm{l}}$.

\subsection*{Polar-distorted B phase}

The lowest temperature phase transition to the polar-distorted B phase (PdB) may in principle occur via a first-order transition from the PdA phase, or via a second-order phase transition from the polar phase. For the experimental conditions studied here, the transition occurs via a first-order phase transition. The order parameter of the PdB phase can be written as
 \begin{equation}
A_{\alpha j} = \sqrt{\frac{1+2q^2}{3}} \Delta_{\mathrm{PdB}} e^{i\phi} (\hat{\bm{d}}_\alpha \hat{\bm{z}}_{j} + q_{1} \hat{\bm{e}}^1_\alpha \hat{\bm{x}}_j 
+ q_{2} \hat{\bm{e}}^2_\alpha \hat{\bm{y}}_j ) \,,
\label{distBop}
\end{equation}
where $\lvert q_1 \rvert , \lvert q_2 \rvert \in (0,1) $, $\lvert q_1 \rvert = \lvert q_2 \rvert \equiv q$ describes the relative gap size in the plane perpendicular to the strands. Vectors $\hat{\bm{e}}^1$ and $\hat{\bm{e}}^2$ are unit vectors in the spin-space. The maximum gap $\Delta_{\mathrm{PdB}}(T,q)$ is achieved along the direction parallel to the strand orientation. For $q=0$, we obtain the order parameter of the polar phase and for $q=1$, we obtain the order parameter of the isotropic B phase. In zero magnetic field the total symmetry group describing the PdB phase can be written as
\begin{equation}
 H_{\mathrm{PdB}} = D_{\infty J} \times T \times \tilde P \,,
\end{equation}
where the notation $J$ refers to the symmetry of the combined rotation of $L$ and $S$ simultaneously. The group $H_{\mathrm{PdB}}$ is again a subgroup of $H_{\mathrm{P}}$, which reflects the fact that the PdB phase can in principle be obtained by the second-order phase transition from the polar phase. 

The homotopy group $\pi_1(G'/H_{\mathrm{PdB}})= \mathbb{Z}\times \mathbb{Z}$ provides the topological stability of phase vortices and combined orbital and spin disclinations, but not of the half-quantum vortices. It is the lack of the last factor $\mathbb{Z}_2$ in the homotopy group which gives rise to the topologically unstable domain wall terminating on HQVs in the PdB phase: the  KLS  wall bounded by HQV strings.


\begin{figure*}
 \includegraphics[width=\linewidth]{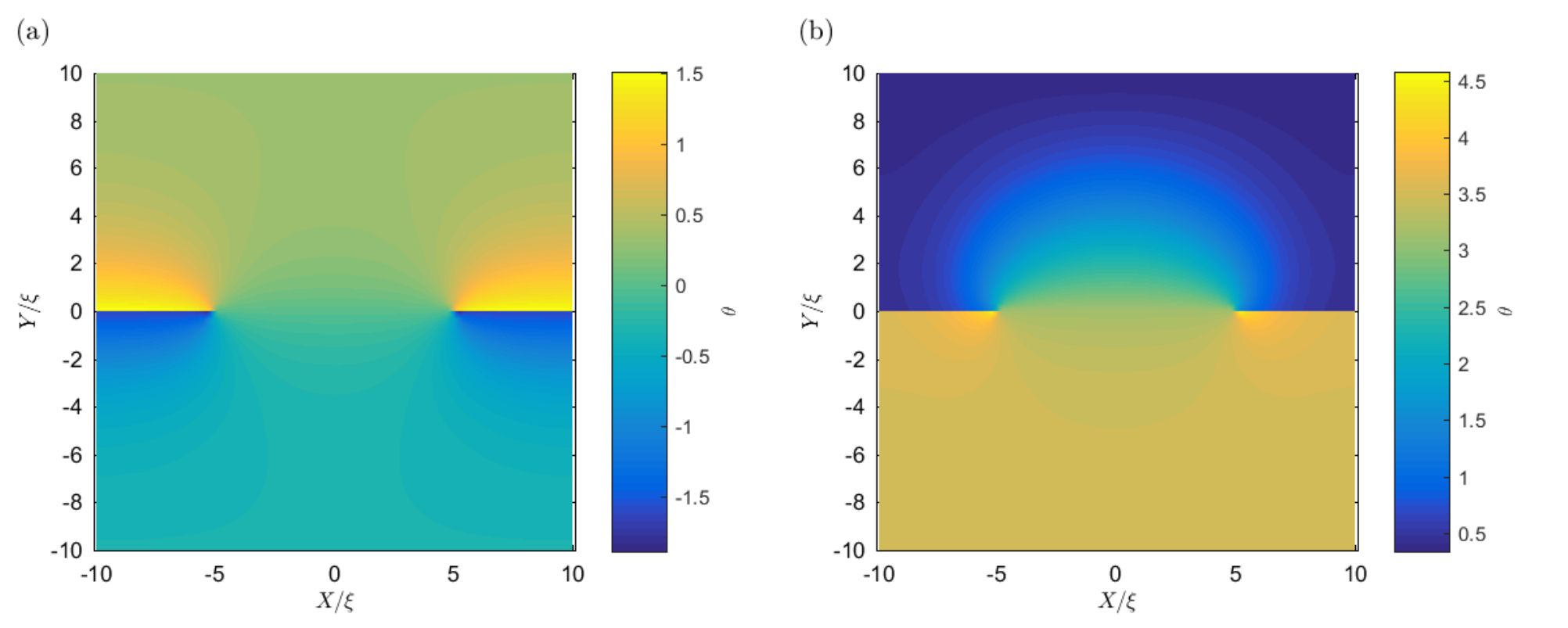}
 \caption{\textbf{Supplementary Note 2 -- Extended data figure for 2D calculations:} 2D numerical calculations of the distribution of angle $\theta$ inside and in the vicinity of KLS walls. In both (a) and (b) panels the upper half corresponds to $q_2 > 0$ and the lower half to $q_2<0$. The  KLS  walls are located on the $y=0$ axis between $X/\xi \in (-5,5)$ and virtual jumps in the order parameter on the same axis between $X/\xi \in [-10,-5]$ and $X/\xi \in [5,10]$. Plot (a) corresponds to the situation where the KLS  walls and $(\pi-2\theta_0)$-solitons are located between different HQV pairs. Plot (b) corresponds to the situation where the KLS  Walls and $(\pi-2\theta_0)$-solitons extend between the same HQV pair. Parameter value $q=0.4$ was used in the calculations.} 
 \label{fig:extended_data}
\end{figure*}

\section*{Supplementary note 3: Free energy of the polar-distorted B phase}

The Landau-Ginzburg free-energy of the PdB phase is given as (summation over repeated indices is assumed)
\begin{align}
F &= \int d^3 x~ (f_{\rm grad}+f_{\rm bulk}+f_{\rm nafen}+f_{\bm{H}}+f_{\rm so}) \label{eq:free-energy}\\
f_{\rm grad} &= \tfrac{K_1}{2} (\nabla_i A_{\mu j}) (\nabla_i A^*_{\mu j}) \nonumber \\ & \quad + \tfrac{K_2 + K_3}{2}(\nabla_{i} A_{\mu i})(\nabla_j A^*_{\mu j}) \\ 
f_{\rm so} &= g_{\rm so} \left(\vert\textrm{Tr}(A) \vert^2 + \textrm{Tr}(A A^*) \right) \label{eq:fso} \\ 
f_{\rm nafen} &= \tfrac{1}{2}\eta_{ij}A_{\mu i}A^{*}_{\mu j}, \quad \eta_{ij} = \eta\delta_{ij}+\Delta \eta \hat{\bm{z}}_i\hat{\bm{z}}_j, \\
f_{\bm{H}} &= -\tfrac{1}{2} \bm{H}\boldsymbol{\chi} \bm{H}, \quad \chi_{\alpha \beta} = \chi_{\rm N}\delta_{\alpha\beta} - \tilde{\alpha}A_{\alpha i}A^*_{\beta i}, 
\end{align}
where $f_{\rm grad}$ is the gradient energy, $f_{\rm bulk}$ is the standard bulk condensation energy of $^3$He \cite{VollhardtWoelfle}, with order parameter matrix $A_{\mu i}$ corresponding to $d_{\mu i}$ in the notation of Ref. \onlinecite{VollhardtWoelfle}. With this convention, the spin-orbit coupling corresponds to $g_{\rm so} = \tfrac{1}{5}\lambda_D N_{\rm F}$ of Ref. \onlinecite{VollhardtWoelfle}. The effect of the nafen confinement $f_{\rm nafen}$, with the uniaxial anisotropy $\Delta \eta$ along $\parallel \hat{\bm{z}}$, is to renormalize the quadratic coefficients $\propto \Delta_{\rm PdB}^2$ since $T_{c, \rm nafen} < T_{c}$ in bulk \cite{DistB}. The magnetic susceptibility tensor $\boldsymbol{\chi}$ and the coefficient $\tilde{\alpha}$ for the PdB phase are also found in Ref. \onlinecite{DistB}.

The order parameter of polar distorted $^3$He-B in nafen is parametrized by Eq.~\eqref{distBop}. Here we concentrate on the limit of large polar anisotropy of the superfluid with the magnetic field $\bm{H} \neq 0$ transverse to the uniaxial anisotropy, i.e. the condition $\vert q_{1,2} \vert \ll 1$ holds. The corresponding ansatz for the spin part of the order parameter is 
\begin{align}
\hat{\bm{d}} &= R(\hat{\bm{y}},\theta) \hat{\bm{x}} =\cos \theta \hat{\bm{x}} - \sin \theta \hat{\bm{z}} \nonumber\\
\hat{\bm{e}}^1 &=-R(\hat{\bm{y}},\theta)\hat{\bm{z}} =-\cos \theta \hat{\bm{z}} - \sin \theta \hat{\bm{x}} \label{eq:theta_ansatz}\\
\hat{\bm{e}}^2 &= \hat{\bm{y}}. \nonumber
\end{align}
We emphasize that with this parametrization $\hat{\bm{e}}^1 \times \hat{\bm{e}}^2=\hat{\bm{d}}$ and the polar phase is obtained by setting $q=0$, whereas the bulk B-phase corresponds to $q_1=\pm q_2=1$. The degeneracy parameters for the bulk B phase are given by the rotation axis $\hat{\bm n} = \hat{\bm y}$ and angle $\cos\theta_{\hat{\bm n}} = \sin \theta$, which is the spin-orbit resolved Leggett angle \cite{VollhardtWoelfle}. To first order in $q_{1,2}$, the gradient energy is simply $f_{\rm grad} = \tfrac{1}{2}K^{\theta}_{ij}(\nabla_i\theta)(\nabla_j\theta)$ with $K^{\theta}_{ij} = K_1(\delta_{ij}-\hat{\bm z}_i\hat{\bm z}_j) + (K_1+K_2+K_3)\hat{\bm z}_i\hat{\bm z}_j$. With the ansatz \eqref{distBop}, the spin-orbit interaction takes the form
\begin{align}
f_{\rm so} &= 2g_{\rm so} \Delta_{\rm PdB}^2 \tilde{f}_{\rm so}(\theta), \label{eq:fso_ansatz} \\ 
\tilde{f}_{\rm so}(\theta) &\equiv (1+q_1)^2\sin^2\theta - (1+q_1)q_2 \sin \theta - q_1+q_2^2. \nonumber 
\end{align}

\section*{Supplementary note 4: Free energy of solitons and domain walls}

As discussed in the main text, the symmetry in the plane transverse to the anisotropy axis is broken by the magnetic field and/or the presence of defects. In equilibrium, $\hat{\bm{d}} \perp \bm{H}$ and we take $\hat{\bm{e}}^2 = \hat{\bm{y}}$ along $\bm{H}$. With this notation the KLS wall is a domain wall in $q_2 $.

We now describe the order-parameter textures in the PdB phase which are associated with the HQVs pinned to the nafen strands. Similar, non-topological defects in the B-phase were already discussed in Refs.~\citenum{Volovik1990} and \citenum{SalomaaVolovik1988}, where the hierarchy of non-topological defects with length scales $\xi, \xi_{\parallel, \perp}$ (GL coherence lengths), $\xi_D$ (dipole length) and $\xi_H$ (magnetic healing length) was emphasized. More recently, non-topological defects with size $\xi_D$ were termed ``soft" and those of size $\xi$ ``hard" in Ref.~\citenum{Thuneberg2014}.

As discussed in the main text, the HQVs in the PdB phase are accompanied with a spin soliton and a KLS domain wall in the superfluid order parameter.

\begin{figure}
\includegraphics[width=1\linewidth]{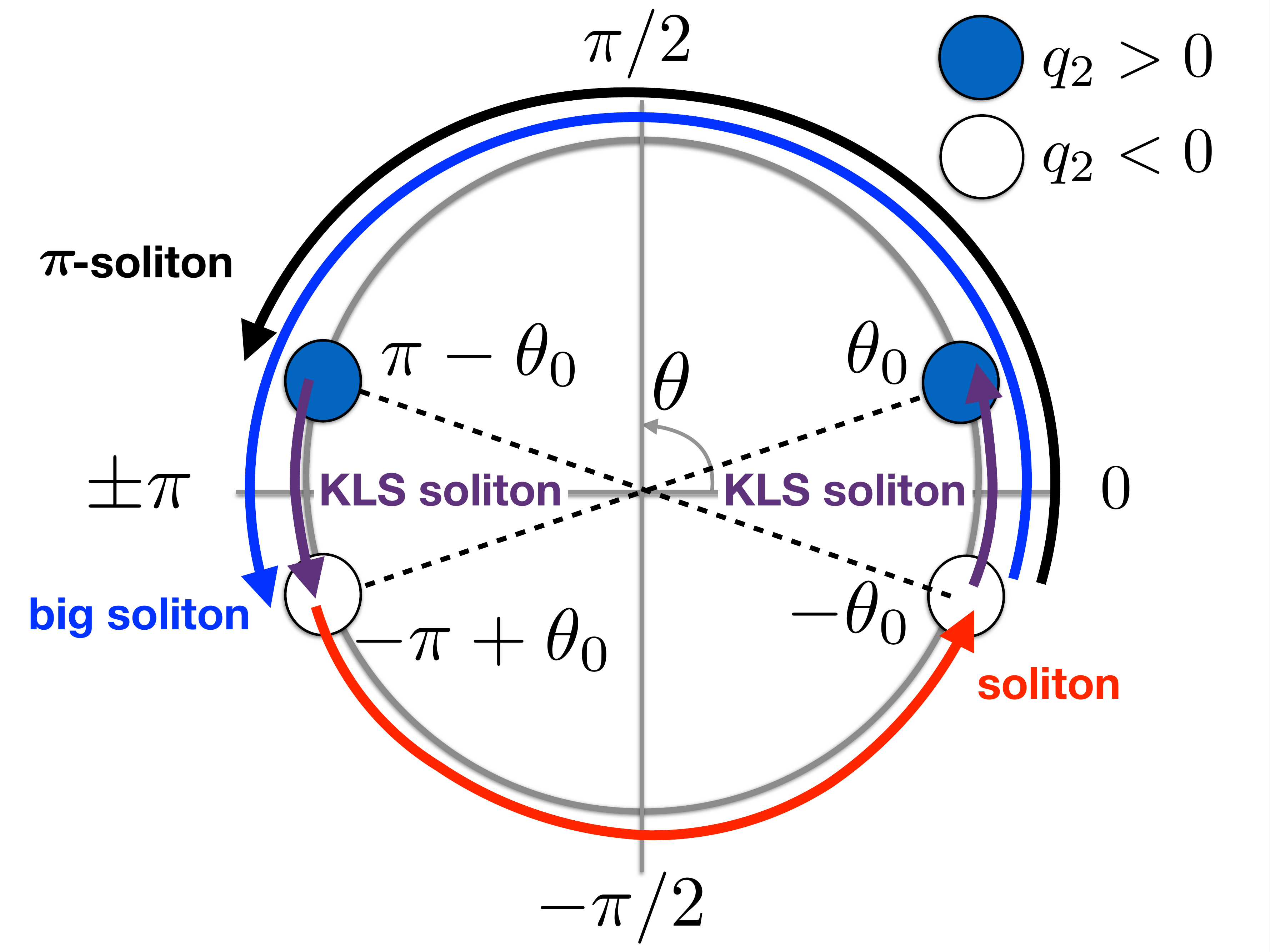}
\caption{ The figure shows schematic illustration of the possible soliton solutions in the PdB phase. The blue circles correspond to the minima in Eq. \eqref{eq:so_minima} for $q_2>0$ and white circles to $q_2<0$. Possible soliton solutions of the spin-winding angle $\theta$ of the order parameter Eq.~\eqref{eq:theta_ansatz} are shown with arrows. The solitons are not symmetric under $\theta\to \theta+\pi$ (``small'' and ``big" solitons) and across the  KLS  wall, $q_2$ changes sign (the ``KLS " and ``$\pi$"-solitons).} 
\label{fig:so_minima}
\end{figure}

\subsection*{KLS  walls}

The KLS domain wall is the change of sign in the transverse, in-plane gap components $q_1 \Delta_{\rm PdB} , q_2 \Delta_{\rm PdB}$ determined by the in-plane coherence length $\xi_{\perp}$. Without loss of generality, we fix the domain wall to act only on $q_2 \Delta_{\rm PdB}$ and the direction normal to the domain wall to be $\hat{\bm{x}}$. Let us write the bulk free energy of the PdB phase in nafen, Eq. \eqref{eq:free-energy}, as the sum of the polar phase free energy and the planar distortion $f_{\rm PdB} = f_{\rm P}+f_{\perp}$, 
\begin{align}
f_{\rm P} &= \alpha_{\parallel}\Delta_{\rm PdB}^2 + \beta_{12345}\Delta_{\rm PdB}^4,\\
f_{\perp} &= ( \alpha_{\perp} + 2\beta_{12}\Delta_{\rm PdB}^2 )(q_{1}^{2} \Delta_{\rm dB}^2+q_{2}^{2} \Delta_{\rm PdB}^2) \nonumber\\
&+ 2\beta_{12} q_{1}^{2}\Delta_{\rm PdB}^2 q_{2}^{2}\Delta_{\rm PdB}^2 \\
&+\beta_{12345} (q_{1}^{4} \Delta_{\rm PdB}^4+q_{2}^{4} \Delta_{\rm PdB}^4), \nonumber\\
f_{\rm grad}[q_{2} \Delta_{\rm PdB}] & = \tfrac{1}{2}K^{(2)}_{ij} (\nabla_i q_{2}\Delta_{\rm PdB})(\nabla_j q_{2} \Delta_{\rm PdB}), 
\end{align}
where $K^{(2)}_{ij} = K_1 (\delta_{ij}-\hat{\bm{y}}_i\hat{\bm{y}}_j) + (K_1+K_2+K_3)\hat{\bm{y}}_i\hat{\bm{y}}_j$. From $f_{\rm P}$ we obtain that $\Delta_{\rm PdB}^2 = -\alpha_{\parallel}/2\beta_{12345}$. For an infinite  KLS  wall along the $y$-axis, the order parameter is given by
\begin{align}
\xi_{\perp2}^2\tfrac{d^2}{dx^2} q_{2} \Delta_{\rm PdB}  = -q_{2} \Delta_{\rm PdB} + \tfrac{(q_{2} \Delta_{\rm PdB})^3}{(q_{2} \Delta_{\rm PdB}^0)^2}, \\
q_{2}(x) \Delta_{\rm PdB} = q_{2} \Delta_{\rm PdB} \tanh \left( \tfrac{x}{\xi_{\perp2}} \right).
\end{align} 
where $\xi_{\perp2}^2/\xi_{\parallel}^2 \sim q^{-2} \gg 1$ and the  KLS  wall thickness is $q^{-1} \xi_{\parallel}$. For the KLS wall to be stable, this should be $\gg \xi_{\parallel}$, i.e. the distortion $q$ should be small.  However, on the length scale of the dipole length, $\xi_D^2 \sim K_1/g_{\rm so}$, relevant in NMR experiments, the  KLS  wall is thin, since $g_{\rm so} \ll -\alpha_{\perp}$. The free energy of the domain wall per unit area is
\begin{align}
\sigma_{\rm KLS} \sim \xi_{\perp2} \Delta f 
\sim -\xi_{\parallel}q f_{\perp},
\end{align}
where $\Delta f \approx f_P-f_{\rm PdB}$. This surface tension makes the isolated HQVs unstable in the PdB phase without the nafen-pinning \cite{Volovik1990,hqv_stability}.

\subsection*{Spin solitons}

Spin solitons have thickness of the order of the dipole length $\xi_D$. The distribution of $\theta(\bm{r})$ in the presence of HQV spin solitons is found as a minimum of energy in Eq.~\eqref{eq:free-energy},
\begin{align}
-\nabla_i\frac{\delta F}{\delta \nabla_i \theta} + \frac{\delta F}{\delta \theta} = -\xi_{D,ij}^2\nabla_i \nabla_j \theta(\bm{r})+\tfrac{1}{2}\frac{\delta \tilde{f}_{\rm so}(\theta(\bm{r}))}{\delta \theta} = 0. \label{eq:theta_EOM}
\end{align}
In bulk the energy is minimized for a homogeneous $\theta = \theta_0$ or $\pi-\theta_0$, where
\begin{align}
\label{eq:so_minima}
\theta_{0} =  \arcsin \frac{q_2}{2(1+q_1)}.
\end{align}
The minima for the spin-orbit potential $f_{\rm so}(\theta)$ depend on the sign of $q_2$, which changes across the  KLS  walls. In contrast to the polar phase with $q_1=q_2 =0$, the potential is no longer symmetric under $\theta \to \theta+\pi$, see Supplementary Figure~\ref{fig:so_minima}.

The equations can be solved analytically for the infinite soliton uniform in the $y$- and $z$-directions. Integrating Eq. \eqref{eq:theta_EOM} over $y$ and $z$ we obtain
\begin{align}
\xi_D^2(\theta')^2 = (1+q_1)^2(\sin \theta -\sin\theta_{0})^2 + C, \label{eq:constant_of_motion}
\end{align}
where $C=0$ by the bulk boundary conditions $\theta(x \to \pm \infty) = \theta_{0}$, $\theta'(x \to \pm\infty)=0$. The soliton solutions are
\begin{align}
\theta(\pm \tilde{x}) &= \mp \pi/2 \mp 2 \arctan f_{\mp}(\tilde{x};s_{0}), \label{eq:soliton_sol}
\end{align}
where we scaled $\tilde{x} \equiv (1+q_1)x/\xi_D$, abbreviated $s_{0} \equiv \sin \theta_{0}$ and
\begin{align}
f_{\mp}(\tilde{x};s_{0}) &=  \sqrt{\tfrac{1\mp s_0}{1\pm s_0}} \tanh(\sqrt{1-s_0^2}\tilde{x}/2 )
\end{align} 
where there are two solutions corresponding to the two signs in Eq.~\eqref{eq:constant_of_motion} and we have used the boundary conditions $\theta(0) = \{-\pi/2,+\pi/2 \}$. The two soliton solutions in Eq. \eqref{eq:soliton_sol} have windings $\pi\mp 2\theta_0$. Clearly the two solutions interchange as $s_0 \to -s_0$.

When $\xi \ll \xi_D$,  we can approximate the  KLS  domain wall as $q_2(x)= q \textrm{sign}(x)$. Across a  KLS  wall, $\theta_{0} \to - \theta_{0}$ and $s_0 \to -s_0$ and we can respectively join the corresponding solutions with boundary conditions $\theta(0) = 0$ or $\pm\pi/2$ at the  KLS  wall, see Supplementary Figures~\ref{fig:so_minima} and \ref{fig:soliton_solutions}. In particular, we can find a solution with $\Delta \theta = 2 \theta_0$ that crosses $\theta(0) =0$ and a solution with $\Delta \theta = \pi$, $\theta(0) = \pm \pi/2$ composed of a small and big soliton on the opposite sides of the domain wall. For $q_2=\textrm{sgn} (x) \vert q_2 \vert $, the  KLS  soliton solution with $\Delta \theta = 2\theta_0$ is given by 
\begin{align}
\theta_{\rm  KLS }(\tilde{x}) = 2 \arctan \left( \tfrac{\textrm{sgn}(x)\vert s_0 \vert }{1+\sqrt{1-s_0^2} \coth(\sqrt{1-s_{0}^2}\vert \tilde{x}\vert/2)} \right). \label{eq:KLS_soliton}
\end{align}
The plots of the soliton solutions interpolating between the PdB spin-orbit energy minima in Supplementary Figure~\ref{fig:so_minima} are found in Supplementary Figure~\ref{fig:soliton_solutions}: In summary, we find two solitons (``soliton" and ``big soliton") without  KLS  wall and two solitons (the `` KLS  soliton" and the ``$\pi$-soliton") connecting the solutions with opposite sign of $s_{0}$. In terms of the more realistic 2D HQV-pair structures depicted in Supplementary Figure~\ref{fig:extended_data} or in the main text, the separate 1D small soliton and  KLS  soliton roughly corresponds to the case shown in  (a) in Supplementary Figure~\ref{fig:extended_data} with the  KLS  wall outside the spin soliton, whereas the $\pi$-soliton corresponds to that shown in Supplementary Figure~\ref{fig:extended_data} (b).

The free energy per unit area of a 1D spin soliton is
\begin{align}
\sigma_{\rm spin} &= \tfrac{1}{L R_{\rm HQV}} \int d^3 \bm{r}~(f[\theta(\bm{r})]-f[\theta_{0}]) \\
&\approx \tfrac{\chi}{2\gamma^2} \Omega_{\rm PdB}^2 \int d x~ 2(\tilde{f}_{\rm so}(\theta(x)) -\tilde{f}_{\rm so}(\theta_0)) \sim \xi_D \Delta f_{\rm soliton} \nonumber.
\end{align}
where $L$ is the linear size along $z$-direction (the height of the sample), and $R_{\rm HQV}$ the the linear size along $y$-direction (the distance between HQVs bounding the soliton).

\begin{figure}
 \includegraphics[width=1\linewidth]{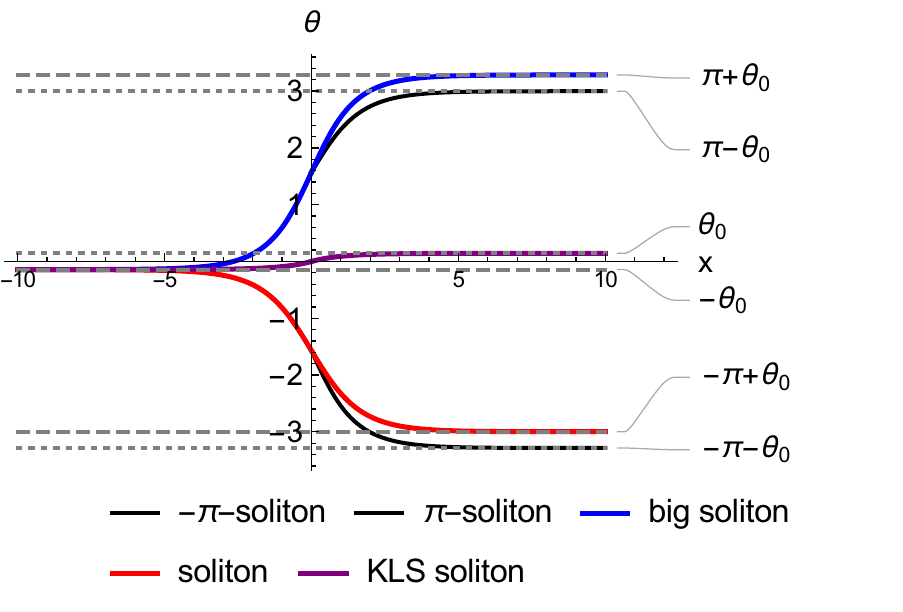}
 \caption{The figure shows the 1D soliton solutions of Eq.~\eqref{eq:theta_EOM} for $q_2(-\infty)<0$. The ordinary soliton has $\Delta \theta = \pi + 2 \vert\theta_{0} \vert$ and $\theta(0)=-\pi/2$. The solution with $\theta(0) = \pi/2$ leads to the big soliton with winding $\Delta \theta = \pi-2 \vert \theta_{0} \vert$. Across the  KLS  wall, one must join the solutions with different signs of $s_0$ with $\Delta \theta = 2\theta_{0}$ or $\pi$. The latter is a composite of a big soliton and an ordinary soliton across the KLS wall.}  
 \label{fig:soliton_solutions}
\end{figure}

\section*{Supplementary note 5: Pinning of HQV by a columnar defect}

Let us consider what happens with HQVs, when the phase transition is crossed between polar phase and the PdB phase. Our experiments demonstrate that if originally the polar phase contains pinned HQVs, they survive the transitions to the PdB phase and back to the polar phase. From this one can conclude that the HQVs remained pinned even after the formation of a KLS domain wall formed between two HQVs, demonstrating that HQVs are so strongly pinned that the tension of the KLS wall can not unpin vortices. Let us consider the pinning in more detail (assuming $\hbar = 1$ and $k_{\mathrm{B}}=1$).

The radius of the columnar defect -- the nafen strand -- is small compared to the coherence length in superfluid $^3$He. According to Ref.~\citenum{0022-3719-10-16-018} the characteristic energy of the order parameter distortion produced by a mesoscopic object of size $R< \xi_0 \equiv \xi(T=0)$ is (per unit length of the cylinder):
\begin{equation}
E_{\mathrm{P}} \sim  k_{\mathrm{F}}^2 R \frac{\Delta^2}{T_\mathrm{c}}   \,\,,\,\, R<\xi_0
 \,,
\label{EnergyPinning}
\end{equation}
where $k_{\mathrm{F}}$ is the Fermi momentum, $\Delta \sim v_{\mathrm{F}} \xi^{-1}$ is the superfluid gap (here we use general gap notation, since this is an order-of magnitude estimation and $\Delta \sim \Delta_{\mathrm{PdB}}$), and $v_{\mathrm{F}}$ is the Fermi velocity. This equation was used in particular for the estimation of the orientational energy of the nafen strands on the orbital vector $\hat{\bf l}$ in $^3$He-A in relation to the Larkin-Imry-Ma effect \cite{Volovik1996}.  The
Larkin-Imry-Ma effect due to the random anisotropy produced by the random orientation of strands was observed later \cite{Dmitriev2010,halperin_nphys}.

Eq.~(\ref{EnergyPinning}) can be applied for the pinning of the texture $\Delta({\bf r})$ by columnar defects -- the nafen strands.  The pinning force comes from the coordinate dependence of the energy  of the  columnar object in the texture: $F_{\mathrm{P}} \sim \nabla E_{\mathrm{P}}$. For textures with characteristic length scale $\xi$, one has $F_{\mathrm{P}}\sim \nabla E_{\mathrm{P}} \sim E_{\mathrm{P}}/\xi$.
For vortices, including  the half-quantum vortices observed in Ref.~\citenum{Autti2016}, the pinning force from the columnar defect of radius $R$ is (assuming $\Delta/T_{\mathrm{c}} \sim \xi_0/\xi$):
\begin{equation}
F_{\mathrm{P}} \sim k_{\mathrm{F}}^2v_{\mathrm{F}}  \frac {R}{\xi^2} \frac{\Delta}{T_{\mathrm{c}}} \sim  k_{\mathrm{F}}^2v_{\mathrm{F}}  \frac {R\xi_0}{\xi^3}  \,\,,\,\, R<\xi_0
 \,.
\label{VortexPinning}
\end{equation}
Let us compare the pinning force with the tension of the KLS wall of thickness $\xi_{\mathrm{W}} \sim q^{-1}\xi \gg \xi$, given by
\begin{equation}
F_{\mathrm{KLS}} \sim k_{\mathrm{F}}^2v_{\mathrm{F}}  \frac {q^{2}}{\xi_{\mathrm{W}}} \frac{\Delta^2}{T_{\mathrm{c}}^2} \sim k_{\mathrm{F}}^2v_{\mathrm{F}} q^3 \frac {\xi_0^2}{\xi^3} 
 \,.
\label{VortexPinning4}
\end{equation}
The tension from the KLS wall can not unpin the HQV if $F_{\mathrm{KLS}}<F_{\mathrm{P}}$, or if
\begin{equation}
q^3 < \frac{R}{\xi_0} < 1
 \,.
\label{VortexPinningKibble}
\end{equation}
Close to the transition from the polar to PdB phase, the HQVs remained pinned, while the KLS wall is pinned by the pinned HQVs.

Let us consider the pinning force by the columnar defect for different ranges of $R$. For $R>\xi$ the pinning does not depend on $R$, but instead is given by the characteristic length scale $\xi$. The dependence of the pinning force on $R$ is given by
\begin{eqnarray}
\frac{F_{\mathrm{P}} }{k_{\mathrm{F}}^2v_{\mathrm{F}}}  \sim   \frac {R\xi_0}{\xi^3} \, \,\,,\,\, R<\xi_0
 \,,
\label{VortexPinning1}
\\
\frac{F_{\mathrm{P}} }{k_{\mathrm{F}}^2v_{\mathrm{F}}}  \sim  \frac {R^2}{\xi^3}  \,\,\,,\,\, \xi_0 < R<\xi
 \,,
\label{VortexPinning2}
\\
\frac{F_{\mathrm{P}} }{k_{\mathrm{F}}^2v_{\mathrm{F}}}  \sim  \frac {1}{\xi}  \,\,\,,\,\, R>\xi
 \,.
\label{VortexPinning3}
\end{eqnarray}

\section*{Supplementary note 6: Spin waves and NMR in the PdB phase}

We study the HQVs and  KLS  domain walls in the PdB phase via their influence on the NMR spin-wave spectrum through the order parameter textures of solitons. The relevant Hamiltonian is given by the magnetic field energy and the superfluid spin degrees of freedom in the London limit,
\begin{align}
\bm{H} &= \frac{1}{2} \gamma^2 \bm{S} \chi^{-1} \bm{S} - \gamma \bm{H}\cdot \bm{S} +  f_{\rm grad}+f_{\rm so}, 
\end{align}
where $\bm{S}$ is the total spin density, $\gamma$ is the gyromagnetic ratio of $^3$He and $\chi$ is the principal axis of the magnetic susceptibility tensor along $\hat{\bm{d}}$. The  Leggett equations for the spin $\bm{S}$ and the order parameter spin-triad $\hat{\bm{e}}^I = \{\hat{\bm{e}}^1, \hat{\bm{e}}^2, \hat{\bm{d}}\}$, where $I=1,2,3$, are
\begin{align}
\doo_t \bm{S} &= \{\bm{S},\bm{H} \}  = \gamma \bm{S} \times \bm{H} + \tfrac{\delta(f_{\rm grad}+f_{\rm so})}{\delta \hat{\bm{e}}^I} \{ \bm{S}, \hat{\bm{e}}^I\}\\
\doo_t \hat{\bm{e}}^I &= \{\hat{\bm{e}}^I,\bm{H}\} = -\tfrac{\gamma^2}{\chi} \hat{\bm e}^{I} \times \delta \bm{S},
\end{align} 
with the semiclassical Poisson brackets $\{S_{\alpha},S_{\beta}\} = \epsilon_{\alpha\beta\gamma}S_{\gamma}$ and $\{S_{\alpha},e^I_{\beta}\} = \epsilon_{\alpha\beta\gamma}e^I_{\gamma}$. In this parametrization, the spin-orbit interaction Eq.~\eqref{eq:fso} takes the form
\begin{align}
f_{\rm so}[\hat{\bm e}^I] &= 2g_{\rm so } \Delta_{\rm PdB}^2 \left( \hat{\bm e}^I \cdot \bm{B}^{IJ} \cdot \hat{\bm e}^J \right) 
\label{eq:so_matrix}
\end{align}
where $\bm{B}^{IJ} = \tfrac{1}{2}(\delta^{IM}\delta^{JN}+\delta^{IN}\delta^{JM})\tilde{\bm{r}}^M\tilde{\bm{r}}^N$ is a matrix in orbital space $\tilde{\bm{r}}^M = \{q_1\unitvec{x},q_2\unitvec{y},\unitvec{z}\}$ defined by the orbital part of the order parameter and summation over repeated spin-triad and orbital indices $I,J = 1,2,3$ and $M,N = 1,2,3$ is implied, respectively.

We look for solutions in small oscillations to linear order around an equilibrium state $\frac{\delta H}{\delta \bm{S}_0}= \frac{\delta H}{\delta \unitvec{e}_0^I} = 0$. Eliminating $\delta\hat{\bm e}^{I}$ from the system of Leggett equations, we arrive to 
\begin{align}
\omega^2\delta \bm{S} = \im\omega \omega_{\mathrm{L}}(\hat{\bm H} \times \delta\bm{S}) +  \Omega_{\rm PdB}^2 \boldsymbol{\Lambda}\cdot \delta \bm{S}, \label{eq:spin_wave}
\end{align}
where we have defined the ``Leggett frequency" of the PdB phase as the quantity  
\begin{align}
\Omega_{\rm PdB}^2 = 4g_{\rm so} \gamma^2\Delta_{\rm PdB}^2/\chi_{\rm PdB}
\end{align}
which we stress is \emph{not} equal to the longitudinal NMR frequency of the PdB phase for $q_1, q_2\neq 0,1$, see below. The matrix $\boldsymbol{\Lambda} = \boldsymbol{\Lambda}^{\rm grad}+\boldsymbol{\Lambda}^{\rm so} $ is defined by 
\begin{align}
\Lambda^{\rm grad}_{\alpha\beta} &= \xi^2_{D,ij}\bigg((\delta_{\alpha\beta}-\hat{\bm d}^{0}_{\alpha} \hat{\bm d}^0_{\beta})\nabla_i\nabla_j 
+ \hat{\bm d}^0_{\alpha}(\nabla_i \nabla_j \hat{\bm d}^0_{\beta}) \nonumber\\
& \quad \quad - \hat{\bm d}^0_{\beta}(\nabla_i \nabla_j \hat{\bm d}^0_{\alpha}) - 2\hat{\bm d}^0_{\beta}(\nabla_i \hat{\bm d}^0_{\alpha})\nabla_j  \bigg) \label{eq:Lambda_grad} \\
\Lambda_{\alpha\beta}^{\rm so} &=  e^I_{0\alpha}B^{IJ}_{\beta\delta}e^J_{0\delta} - \tilde{e}^I_{0\nu}B^{IJ}_{\nu\delta}e^J_{0\delta}\delta_{\alpha\beta} +\epsilon_{\alpha\nu\gamma}\epsilon_{\delta\mu\beta} e^I_{0\gamma} B^{IJ}_{\nu\delta}e^J_{0\mu}, \nonumber\\ 
\end{align}
where $B^{IJ}_{\alpha\beta}$ is the matrix in Eq. \eqref{eq:so_matrix} for each $I,J$, $\xi_{D,ij}^2 = K_{ij}^{\theta}/4g_{\rm so}$ and the gradient energy is taken to first order in $q_1,q_2$. In the limit $q_1=q_2=0$, the equations are those of the polar phase \cite{SlavaPolar}; in particular the lowest order $\boldsymbol{\Lambda}^{\rm grad}$ in Eq. \eqref{eq:Lambda_grad} coincides with the expression for the polar phase given in Ref.~\citenum{PhysRevLett.115.165304}. 

With $\bm{H} \parallel \hat{\bm y}$ and within the approximation $\omega \approx \omega_{\mathrm{L}}$ to lowest order in $\frac{\Omega_{\rm PdB}}{\omega_{\rm L}}\ll 1$, Eq. \eqref{eq:spin_wave} separates for transverse $\delta S_{+} = (\delta S_z + \im \delta S_x)/\sqrt{2}$ and longitudinal spin waves as \cite{SlavaPolar}
\begin{align}
\tfrac{\omega^2-\omega_L^2}{\Omega_{\rm PdB}^2}\delta S_+ &= (\Lambda_{xx}+\Lambda_{zz})\delta S_+ + \im(\Lambda_{xz}-\Lambda_{zx})\delta S_+, \\
\tfrac{\omega^2}{\Omega_{\rm PdB}^2} \delta S_{y} &= \Lambda_{yy} \delta S_y.
\end{align}

\subsection*{Transverse magnetic field}

For the transverse orientation of the magnetic field to the uniaxial nafen anisotropy along $\hat{\bm z}$, the order parameter is given by Eq.~\eqref{distBop}. The transverse spin wave equation becomes
\begin{align}
-\tfrac{\omega^2-\omega_{\rm L}^2}{\Omega_{\rm PdB}^2} \Psi_+ = \xi^2_{D,ij}\bigg(\nabla_i\nabla_j +(\nabla_i \theta)(\nabla_j\theta) \bigg)\Psi_+ \label{eq:transverse}\\
+\bigg(\tilde{f}_{\rm so}(\theta) - \tfrac{3}{2}(1+q_1)q_2 \sin\theta + q_2^2 \bigg)\Psi_+ \nonumber
\end{align}
where $\Psi_{+} = e^{\im \theta} \delta S_{+}$, the dimensionless spin-orbit interaction $\tilde{f}_{\rm so}(\theta)$ is defined in Eq. \eqref{eq:fso_ansatz} and the longitudinal spin wave equation is
\begin{align}
-\tfrac{\omega^2}{\Omega_{\rm PdB}^2}\delta S_y  = \xi_{D,ij}^2 \nabla_i\nabla_j\delta S_{y} - \bigg(\frac{\partial^2 \tilde{f}_{\rm so}}{\partial \theta^2} \bigg)\delta S_y. \label{eq:longitudinal}
\end{align}
The transverse frequency shift with uniform $\theta  = \theta_{0}$ (i.e. the response of the bulk) is given as
\begin{align} \label{eq:freq_pdb_tra}
\frac{\omega_{\perp}^2 -\omega_{\rm L}^2}{\Omega_{\rm PdB}^2} = q_1-q_2^2. 
\end{align}
This frequency shift was reported also in Ref.~\citenum{DistB}.

\subsection*{Axial field}

In axial field, i.e. $\bm{H}$ along the uniaxial anistropy, the order parameter is no longer given by Eq. \eqref{distBop}. Since the magnetic field energy is dominating and $\unitvec{d} \perp \bm{H} \parallel \unitvec{y}$ which is also the direction of the uniaxial anisotropy (and not along $\unitvec{z}$ as in the preceding sections), we parametrize the orbital part of order parameter as
\begin{align}
A_{\mu i} = \Delta_{\rm PdB} e^{i \phi} \left( \hat{\bm d}_{\mu} \hat{\bm y}_i + q_1 \hat{\bm e}^{1}_{\mu} \hat{\bm x}_i + q_2 \hat{\bm e}^{2}_{\mu} \hat{\bm z}_i \right), 
\end{align}
whereas the spin part is still given by Eq. \eqref{eq:theta_ansatz} with the rotation angle in the plane perpendicular to the magnetic field. The spin-orbit interaction takes the form $\tilde{f}_{\rm so, \parallel}(\theta) = q_1^2 \sin^2 \theta - q_2 \sin \theta$ with $\sin \theta_{0, \parallel} = 0$ for $\vert q\vert <1/2$ \cite{DistB}. The longitudinal spin wave equation in axial field follows from Eq.~\eqref{eq:longitudinal} with this replacement.

\begin{figure}
 \includegraphics[width=\linewidth]{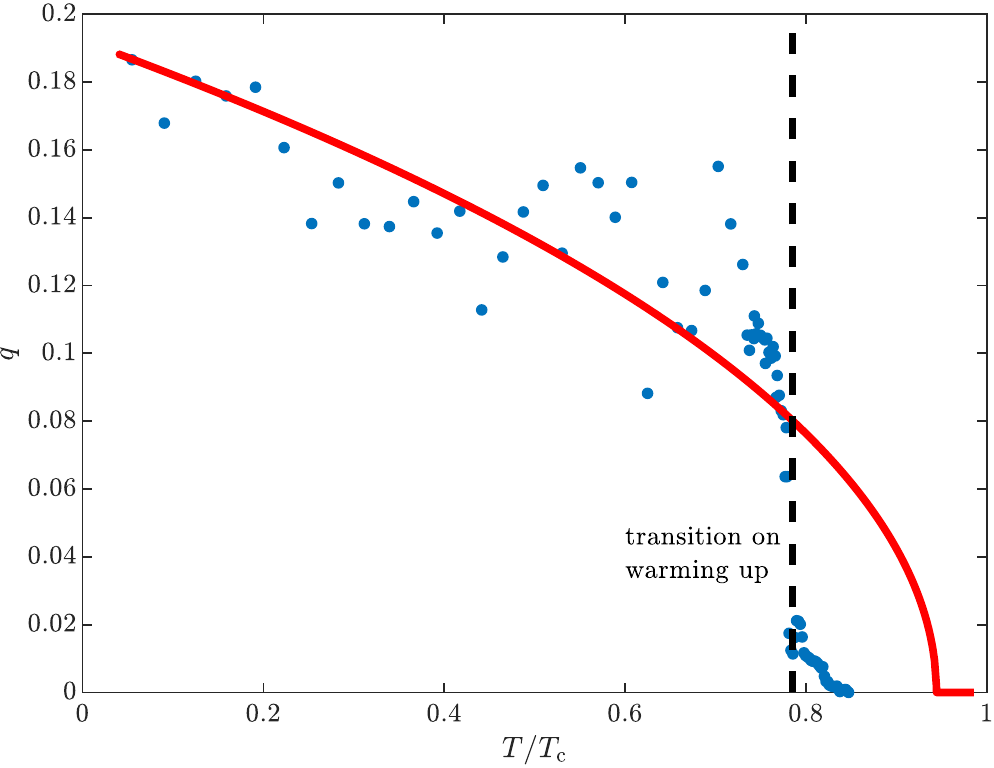}
 \caption{The dots represent the measured values for $q$. The solid red line is an estimation of $q$, calculated based on Ginzburg-Landau theory with strong-coupling corrections using two fitting parameters in the spirit of Ref.~\citenum{DistB} and taking $\beta$-parameter values from Ref.~\citenum{GLstrong}. The PdB phase critical temperature is shown for warming transition to the PdA phase.} 
 \label{gap_ratio}
\end{figure}

The transverse spin wave $\Psi_{+} = e^{\im \theta}\delta S_{+}$ equations are given as
\begin{align}
-\tfrac{\omega_{\parallel}^2-\omega_{\rm L}^2}{\Omega_{\rm PdB}^2} \Psi_+ = \xi^2_{D,ij}\bigg(\nabla_i\nabla_j +(\nabla_i \theta)(\nabla_j\theta) \bigg)\Psi_+ \nonumber\\
+\bigg(-1-\tfrac{5}{2}q_2\sin\theta+q_1^2 \sin^2\theta-q_2^2 \bigg)\Psi_+ . \label{eq:axial}
\end{align}
The homogeneous transverse frequency shift in axial field with uniform $\theta  = \theta_{0, \parallel} = \textrm{sgn}(q_2) \pi/2$ is given as
\begin{align} \label{eq:freq_pdb_ax}
\frac{\omega_{\parallel}^2 -\omega_{\rm L}^2}{\Omega_{\rm PdB}^2} = 1+\frac{5}{2} \vert q_2 \vert ,
\end{align}
which is equal to the value reported in Ref.~\citenum{DistB}.

\section*{Supplementary note 7: Determination of the distortion parameter $q$}

The $q$-parameter value is determined from the frequency shifts in Eqs.\eqref{eq:freq_pdb_tra} and \eqref{eq:freq_pdb_ax}, following a method described in Ref.~\citenum{DistB}. In the experimental region of interest, the distortion factor $q = q_{1} = \vert q_2 \vert$ is
\begin{equation}
 q = \frac{2-5C}{4} - \frac{1}{4}\sqrt{25C^{2} - 36C + 4}, \label{eq:qeq}
\end{equation}
where $C = (\omega_{\perp} - \omega_{\mathrm{L}})/(\omega_{\parallel} - \omega_{\mathrm{L}})$. The expression~\eqref{eq:qeq} is valid in the range $q \in [0,(\sqrt{14}-2)/5]$. We carefully prepare the state by cooling the sample through the superfluid transition temperature in zero rotation in transverse magnetic field to avoid creation of half-quantum vortices. Then we cool the sample down to the lowest temperatures and start warming it up slowly, continuously monitoring the NMR resonance spectrum either in axial or transverse field. This way we can measure the $q$-parameter in the coexistence region of the PdA and PdB phases. The results of our measurements are shown in Supplementary Figure~\ref{gap_ratio}.

\section*{Supplementary note 8: Spin waves on 1D solitons and KLS Walls}

\begin{figure}
\includegraphics[width=\linewidth]{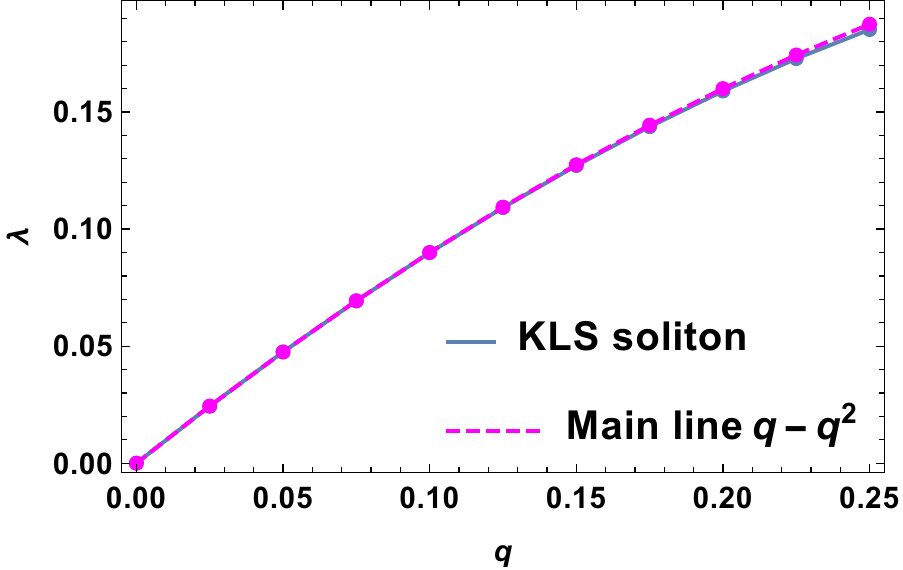}

\vspace*{.25cm}

\includegraphics[width=\linewidth]{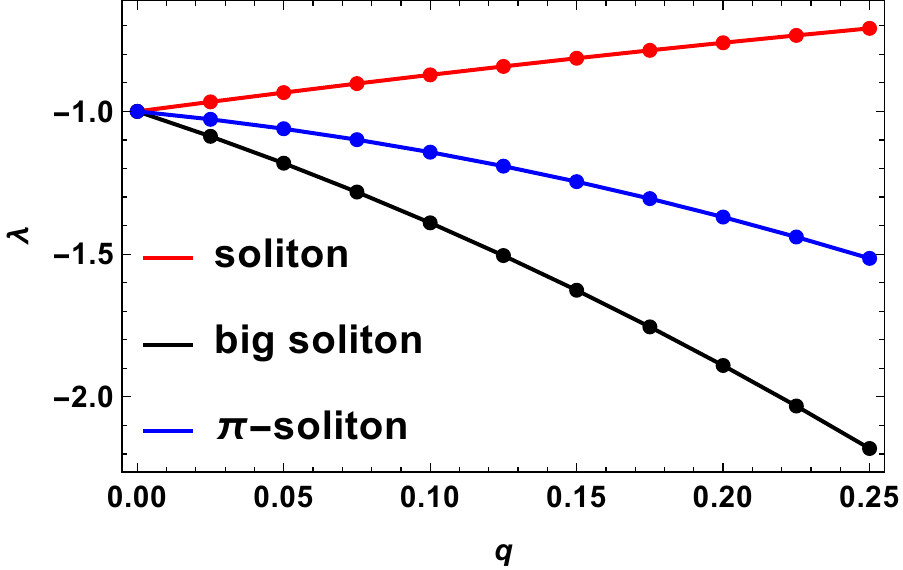}
 \caption{The figures show the NMR resonance eigenvalue $\lambda(q)$ for spin waves localized on infinite 1D solitons. The frequency shift related the the KLS wall (upper figure) is indistinguishable from the frequency shift of the main line in the experimental range of $q$. The lower figure shows the frequency shifts for the other possible solitons as a function of $q$.} 
 \label{fig:lambda_solitons}
\end{figure}

The transverse spin wave equation Eq. \eqref{eq:transverse} takes the form of an eigenvalue equation 
\begin{align}
\lambda \Psi_+ = -\nabla_i\nabla_j \Psi_+ - U(\theta(\bm{r})) \Psi_+ \label{eq:lambda_def}
\end{align} 
with eigenvalue $\lambda \equiv \tfrac{\omega^2-\omega_{\rm L}^2}{\Omega_{\rm PdB}^2}$ and potential 
\begin{align}
U(\theta) = -\xi_{D,ij}\nabla_i \theta\nabla_j\theta -\tilde{f}_{\rm so}(\theta) + \tfrac{3}{2}(1+q_1)q_2 \sin\theta - q_2^2.
\end{align}
The spin-wave spectrum is therefore defined with respect to the order parameter texture $\theta(\bm{r})$, as determined by the GL equations in Eq. \eqref{eq:theta_EOM}. The homogeneous $\delta S_+$ excitation (the main NMR line) is shifted from the Larmor value by $\Delta \omega \equiv \omega-\omega_{\rm L} \approx \tfrac{\Omega_{\rm PdB}^2}{2\omega_{\rm L}}\lambda_{\theta_0}$ with $\lambda_{\theta_0} = U(\theta_{0})$. As discussed, the solutions $\theta(x)$ are analytically tractable in 1D and the spin wave Eq. \eqref{eq:lambda_def} can be efficiently solved numerically.

\begin{figure}

\includegraphics[width=\linewidth]{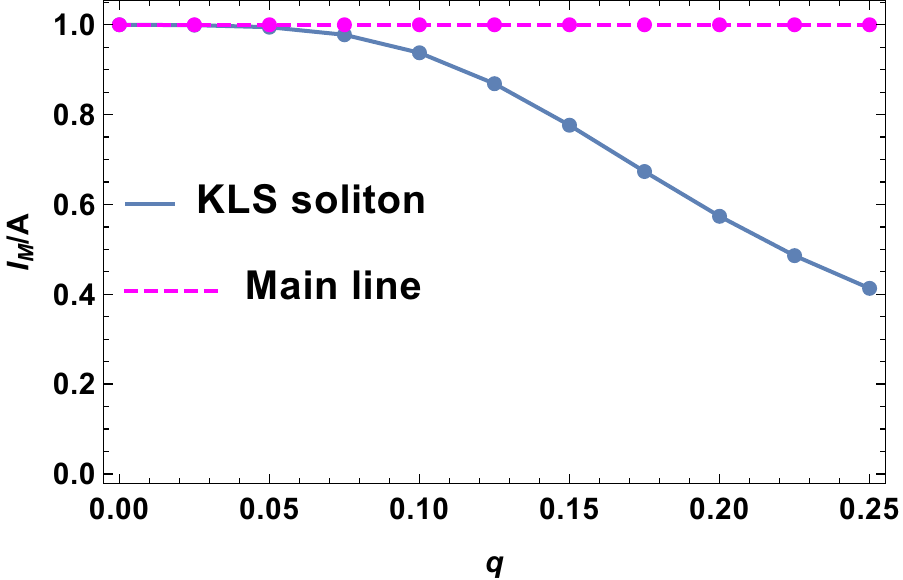}

\vspace*{.25cm}

 \includegraphics[width=\linewidth]{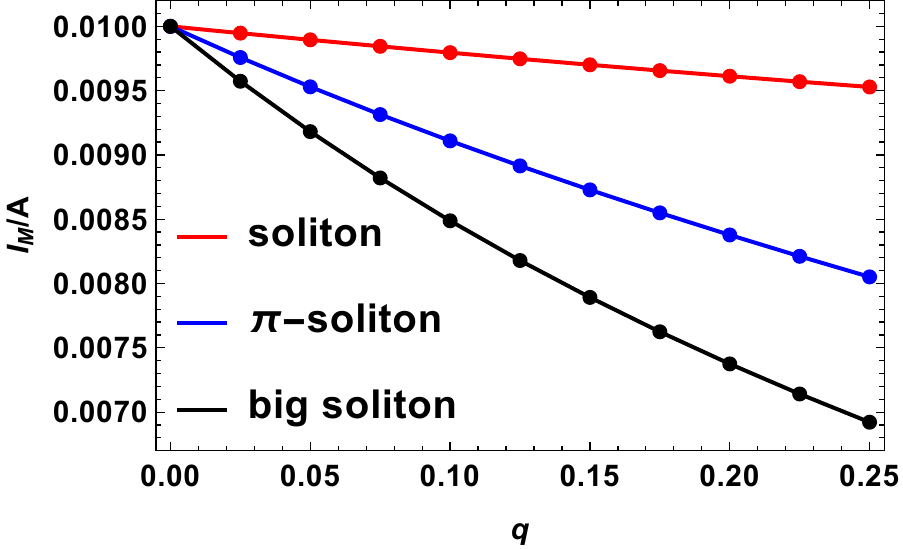}

 \caption{The figures show the NMR oscillator intensities for spin waves on infinite 1D solitons. All solutions show decrease in the oscillator intensity, which results in the decrease of NMR satellite intensity -- as observed in the experiments. However, the observed decrease in the intensity is much larger than the calculated decrease for realistic values of~$q$.}
 \label{fig:osc_strengths}
\end{figure}

Now we want to compute $\lambda$ in the presence of the solitons and  KLS  walls accompanying the HQVs in the PdB phase. The soliton solutions in 1D have an analytical form and the potential is written in terms of $\theta(x)$ that satisfies the equations of motion Eq. \eqref{eq:constant_of_motion} as 
\begin{equation}
U(\theta) = -2 \tilde{f}_{\rm so}(\theta) + \tfrac{3}{2}(1-q_1)q_2 \sin\theta - q_1 - \tfrac{1}{4}q_2^2.
\end{equation}
The eigenvalue equation 
\begin{equation}
\Psi_{+}'' (x)+ U(\theta)\Psi_{+}(x) = \lambda \Psi_{+}(x)
\end{equation}
can be numerically solved for the lowest lying eigenvalue for the spin wave $\Psi_{+}(x)$ localized on an infinite 1D soliton. We obtain eigenvalues $\lambda(q)$ in Supplementary Figure~\ref{fig:lambda_solitons} for the NMR satellite peaks with the relative NMR shifts $\Delta \omega = \lambda(q) \Omega_{\rm PdB}^2/(2\omega_{\mathrm{L}})$. The fit to the temperature dependence of $q(T/T_c)$ in Supplementary Figure~\ref{gap_ratio} then leads to the values $\lambda(T/T_c)$ shown in the main text.

The experimentally measured NMR intensity relative to the bulk is given as (where the total magnetization per mode is $I_M$) \cite{SalomaaVolovik85, RevModPhys.59.533, HuMaki87}
\begin{equation}
I_M / I_0\propto \tfrac{1}{2}n_{\rm HQV} \frac{\vert \int dA~ \delta S_+(x,y) \vert^2}{\int dA~\vert \delta S_+(x,y) \vert^2} 
\end{equation}
where $\delta S_{+}(x,y) = e^{-\im \theta(x)}\Psi_{+}(x)$ is the physical transverse spin wave excitation and $n_{\rm HQV}$ is the areal density of HQVs. These NMR oscillator intensities are shown in Supplementary Figure~\ref{fig:osc_strengths}.

\end{document}